\documentclass{article}

\usepackage[utf8]{inputenc}
\usepackage{amsmath}
\usepackage{amsthm}
\usepackage{amsfonts}
\usepackage{amscd}
\usepackage{amssymb}
\usepackage{natbib}
\usepackage{url}
\usepackage[table,xcdraw,usenames]{xcolor}

\usepackage{graphicx}
\usepackage{subcaption}
\usepackage{mathtools}
\usepackage{enumitem}
\usepackage{bm}
\usepackage{comment}
\usepackage{pdfpages}

\usepackage{hyperref}
\usepackage{caption}
\usepackage{float}
\usepackage{tikz}
\usepackage{multirow}
\usepackage[linesnumbered, ruled,vlined]{algorithm2e}
\usepackage{pdflscape}
\usepackage{etoolbox}

\usepackage{geometry}
\newcommand{\weight}{\pi}

\newcommand{\y}{\textbf{y}}

\newcommand{\x}{\textbf{v}}

\newcommand{\V}{\textbf{V}}

\newcommand{\simiid}{\stackrel{iid}{\sim}} 
\newcommand{\indep}{\perp \!\!\! \perp } 
\def\mrm#1{\mathrm{#1}} 
\def\mbi#1{\boldsymbol{#1}} 
\def\v#1{\mbi{#1}} 
\def\mc#1{\mathcal{#1}} 
\DeclareMathOperator*{\argmin}{arg\,min} 
\def\E{\mathbb{E}} 
\def\mc#1{\mathcal{#1}}

\newtheorem{lem}{Lemma}

\newtheorem{prop}{Proposition}
\theoremstyle{definition}

\hypersetup{
  linkcolor  = blue,
  citecolor  = blue,
  urlcolor   = blue,
  colorlinks = true,
} 

\newenvironment{proof-of-proposition}[1][{}]{\noindent{\bf
    Proof of Proposition {#1}}
  \hspace*{.5em}}{\qed\bigskip\\}
  \newenvironment{proof-of-corollary}[1][{}]{\noindent{\bf
    Proof of Corollary {#1}}
  \hspace*{.5em}}{\qed\bigskip\\}
\newenvironment{proof-of-lemma}[1][{}]{\noindent{\bf
Proof of Lemma {#1}}
\hspace*{.5em}}{\qed\bigskip\\}

\allowdisplaybreaks

\title{Volatility Forecasting Using Similarity-based Parameter Correction and Aggregated Shock Information}
\author{David P. Lundquist\thanks{davidl11@ilinois.edu}, Daniel J. Eck\thanks{dje13@illinois.edu}\\Department of Statistics, University of Illinois at Urbana-Champaign }

\date{\today}
 
\begin{document}

\maketitle

\begin{abstract}
  We develop a procedure for forecasting the volatility of a time series immediately following a news shock.  Adapting the similarity-based framework of \citet{lin2021minimizing}, we exploit series that have experienced similar shocks.  We aggregate their shock-induced excess volatilities by positing the shocks to be affine functions of exogenous covariates.  The volatility shocks are modeled as random effects and estimated as fixed effects.  The aggregation of these estimates is done in service of adjusting the $h$-step-ahead GARCH forecast of the time series under study by an additive term.  The adjusted and unadjusted forecasts are evaluated using the unobservable but easily-estimated realized volatility (RV).  A real-world application is provided, as are simulation results suggesting the conditions and hyperparameters under which our method thrives.
\end{abstract}

\section{Introduction}

Reacting to a seemingly unprecedented event might involve the question: what, if anything, does it resemble from the past?  Such might be the case with event-driven investing strategies, where the identification of the event could arise via the news pages or corporate communications and hence contains a qualitative, narrative element \citep{Kenton}.  Matching a current crisis to past events is a problem with unsurprising statistical angles: identification, sample size, weighting, risk, and robustness, among many others.  

In the context of foreign exchange rate market structure, \citet{dominguez2006defines} speculate ``[w]hether news is scheduled or non-scheduled its influence on exchange rates may be related
to the state of the market at the time of the news arrival.  News that arrives during periods of
high uncertainty may have different effects on the exchange rate, than news that arrives in
calmer periods." The authors also note that non-scheduled news may require more time for markets to digest, leading to greater heterogeneity (including but not limited to higher dispersion) of responses.  We take inspiration from these arguments, developing a method suitable to the conditions that typically accompany news shocks.  Central to our endeavor is the observation that news shocks have both a qualitative (the description of the event as well as future contingents surrounding it) and a statistical incarnation, i.e. the reaction of quantitative phenomena in response to the qualitative information.

In this work we focus on the second central moment of a time series, the volatility.  One of the most important stochastic phenomena of positively-valued time series $(P_{t})_{t\in\mathbb{N}}$, especially financial time series, is the volatility of the return series $(r_{t})_{t\in\mathbb{N}}$.  A financial asset's price series may exhibit behavior that makes inapplicable and uninterpretable the traditional methods of time series analysis.  In contrast, the return series is scale-free \citep{tsay2005analysis}, easily-interpreted, and often at least weakly stationary.  Our reasons for studying volatility are two-fold.  Whereas once returns were thought to react to specific news events, now stock price movements are believed to be overwhelmingly noise-based (see \citet{boudoukh2019information} and references therein), and even if one could construct credible models for describing and forecasting price series and return series, that would not necessarily tell us much about the variability of such forecasts nor enlighten us about the evolution of the variability of $(P_{t})_{t\in\mathbb{N}}$ and $(r_{t})_{t\in\mathbb{N}}$ or under rapidly changing conditions. 

No matter how a time series or its transformations are modeled, forecasting in the presence of news shocks requires a methodological framework that sensibly incorporates relevant information that has yet to manifest in market price or derivative quantities like volatility.  In this setting, regime-change models (see \citet{bauwens2006regime} and citations therein) are of little use because under the assumption of a known exogenous shock, there is no need to estimate a regime-change time, nor is there data following the exogenous shock event to fit a model.  Asymmetric GARCH models were an early attempt to account for fact that negative returns typically beget greater volatility than positive returns \citep{hansen2012realized}.  Asymmetry and mixed-frequency are employed in \cite{wang2020forecasting} to forecast under the presence of extreme shocks.  Problematically, any asymmetric model will depend upon the observation of a negative return to provide the most updated volatility forecast, but under the circumstances posited herein, no such return has been observed.

Similar problems and more exist for Realized GARCH models \citep{hansen2012realized}, which incorporates observable measures of volatility, known as  ``realized measures", like implied volatility (IV).  Under the assumptions herein, no post-shock data is available, and even if it were, Realized GARCH does not promise to perform well, since Black-Scholes implied volatility is a biased estimator of volatility \citep{mayhew1995implied, christensen1998relation}, with the bias increasing in times of crises, when options may be far out-of-the-money.  GARCH models have been shown slow to adapt to spikes in volatility \citep{andersen2003modeling}.

The approach herein sidesteps the functional complexity posited by Realized GARCH, with its minimum nine parameters to estimate \citep{sharma2016forecasting}, by substituting modeling assumptions.  The method herein proceeds under the assumption that similar news events occasion volatility shocks arising from a common conditional shock distribution.  The procedure proposed does not require post-shock information like returns or market-implied quantities from the time series under study.  Hence, we also avoid questions about what realized measure to use and when as well as questions about the usefulness of high-frequency data, although these remain intriguing avenues for future work.

The primary methodological tool presented in this work is fixed effect estimation followed by a distance-based weighting procedure for pooling those estimates.  The use of fixed effect estimation for the study of structural shocks has a pedigree in macroeconomic analysis (\citet{romer1989does} cited in \citet{kilian2017structural}; see also discussion of determinstic exogenous events in \citet{engle2001good}).  We employ fixed effect estimation on the basis of a well-established conceptual assumption that shocks of economic time series can be modeled as mixtures, in particular, mixtures of ordinary innovations and shocks due to rare events (see \cite{phillips1996forecasting} and references therein).  In the forecasting literature, the term ``intercept correction'' has come to refer to a modeling technique in which nonzero errors are explicitly permitted \citep{hendry1994theory, clements1998forecasting}.  They summarize the literature as distinguishing two families of intercept correction: so-called ``discretionary" intercept corrections that attempt to account for future events, without hard-coding an ad-hoc adjustment into the model specification, and second, ``automated" intercept corrections that attempt to adjust for persistent misspecification using past errors.  \citet{guerron2017macroeconomic} use weighted subsets of a scalar series' own past to correct forecast errors by an additive term. \citet{dendramis2020similarity} a introduces a similarity-based forecasting procedure for time-varying coefficients of a linear model. \citet{foroni2022forecasting} employ a form of intercept correction in order to adjust forecasts to the COVID-19 shock in the spring of 2020 based on the proportional misses of the same model applied to the Great Recession.

A researcher interested in forecast adjustment can choose between procedures that discretionary or automated, a variety of choices for the collection of data assembled to perform the correction, whether the data is internal (i.e. from the time series itself) or external, the parametric term to be corrected (e.g. intercept, coefficients), if any, as well as the correction function (i.e. the mapping from the data to the corrective term), including the weighting applied to the assembled data (e.g. Nearest-Neighbor, arithmetic mean, kernel methods).

Our procedure is a discretionary procedure for intercept correction that  incorporates systematically data internal or external to the time series under study.  The correction function, as we shall see, includes a convex combination of fixed effect estimates from a donor pool.  In  the causal inference literature, a donor pool is a set of units of observation that share distributional properties with the prime object of analysis.  In \citet{abadie2010synthetic}, the authors build upon previous work in causal inference whereby a treatment effect can be estimated via comparison with a synthetic time series that represents the control.  In their setting, the synthetic unit is constructed using a convex combination of the donors.  The particular convex combination employed is a function of the distance between the time series under study and the donors.  Donors that are closer to the time series under study are, \textit{ceteris paribus}, accorded greater weight.  Conversely, donors that are unlike the time series under study receive little to no weight.  \citet{lin2021minimizing} adapt the notion of a donor pool as well as distance-based weighting for the purpose of prediction.  Their one-step-ahead forecasts use distance-based-weighting to aggregate shock estimates from the donor pool's time series according to the similarility to the time series under study.  Their approach does not take into account the ARCH effects commonly observed in time series, especially financial times series, leaving unaccounted for the variability in the variability of time series forecasts.  Outside of \citet{lin2021minimizing}, we know of no prior work that both introduces a parametric specification for nonzero errors and introduces a procedure for weighting appropriately the nonzero errors of similar shocks occurring outside the time series under study.  Likewise, we are not familiar with any prior work that attempts to account for anticipated nonzero errors using an explicit parametric adjustment, i.e., what we will call a ``correction function''.  

The method proposed herein is supported by both numerical evidence as well as a real data example.  The numerical evidence takes the form of simulations and analysis of the method's performance across a considerable grid of varying parameters, allowing us to evaluate reasonable hypotheses about the method but also uncover some intriguing nuances.  The real data example, on the other hand, demonstrates the usefulness of the method by applying it to the most closely-followed democratic political event on earth, the US Presidential Election.  We find that the method would have allowed a researcher to predict the volatility of a financial asset that immediately followed the 2016 US Presidential Election in the United States.  Our discussion of this example sheds light on some subtleties of the method as well as its robustness to various choices that must be made to construct and assemble donors and appropriate covariates.

\section{Setting}\label{Setting}

\begin{figure}[h!]
    \begin{center}
      \begin{tikzpicture}
        \node[anchor=south west,inner sep=0] at (0,0) {\includegraphics[width=\textwidth]{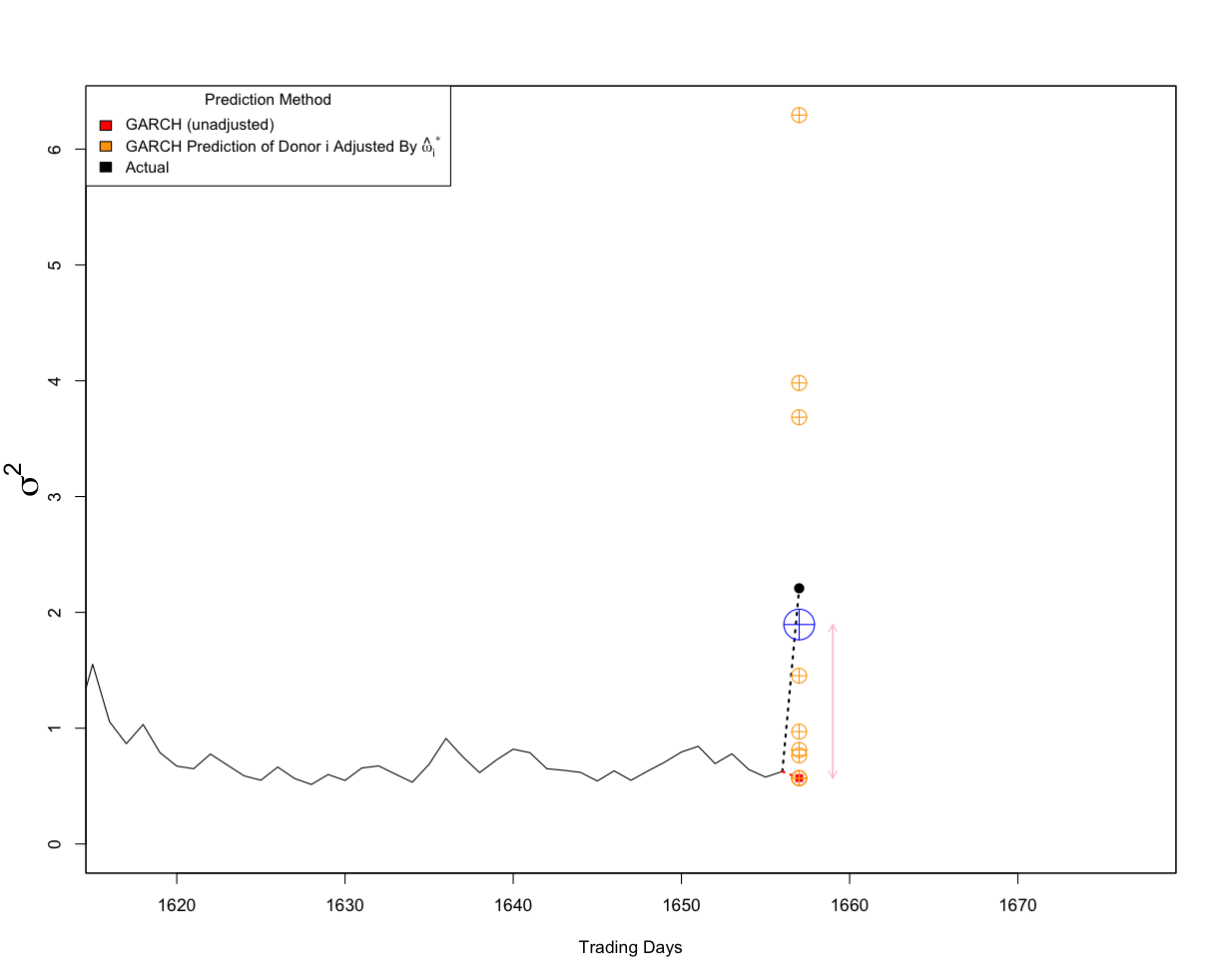}};
        
    \end{tikzpicture}
      \caption{The simulated time series experiences a volatility shock between trading days 1,656 and 1,657.  The GARCH prediction, in red, fails even to approach the volatility spike at $T^{*}+1$, as do several adjusted predictions, which are orange.  In contrast, the GARCH forecast adjusted by $\hat\omega^{*} = \sum^{n+1}_{i=2}\pi_{i}\hat\omega^{*}_{i} $, a convex combination of the estimated shocks in the donor pool, achieves directional correctness as well as a smaller absolute loss in its prediction.  The pink vertical line serves to indicate the adjustment of size $\hat\omega^{*}$ that allows the blue bullseye to more closely approach the ground truth.}    \label{fig:motivating_piece_convex_combination}
   
      \end{center}
    \end{figure}

In order to motivate our procedure, we provide a visual illustration.  In Figure \ref{fig:motivating_piece_convex_combination}, we show how the aggregation of estimated shock-induced excess volatilities from donors in the donor pool works when the correction function is a specially-chosen convex combination of fixed effects from the donor pool.  Our method assumes a credible, parsimonious parameterization in which the shock is an affine transformation of several key covariates.  The key intuition behind this shock parameterization is that as the strength of the linear signal increases relative to the idiosyncratic error, the GARCH estimation of these effects increases in accuracy.  From this, it follows that the aggregated shock estimate increases in accuracy.

  \subsection{A Primer on GARCH}
We define the log return of an asset between $t-1$ and $t$ as $r_{t} = \text{log}(\frac{P_{t}}{P_{t-1}})$, where $P_{t}$ denotes the price at time $t$.  The class of ARIMA($p,d,q$) models  \citep{box2013box} provides a framework for modeling the autoregressive structure of $r_{t}$, all under the umbrella of frequentist statistics.  These models assume a certain dependence structure between $r_{t}$ and $(r_{k})_{k\leq t}$, yet their errors --- often called innovations in the financial time series context due to how they represent the impact of new information --- are nevertheless assumed to be i.i.d. with mean zero and constant variance.  The ARCH \citep{engle1982autoregressive} and GARCH \citep{bollerslev1986generalized} models provide elegant alternatives to the homoskedasticity assumption.  In fact, the GARCH framework in its most basic form disregards $r_{t}$ and instead turns its interest to the series $r_{t}^{2}$ (once properly centered, i.e. after assuming a mean-model for returns).  

To that end, let $a_{t} = r_{t} - \mu_{t}$, where $\mu_{t}$ is the mean of the log return series $r_{t}$.  We thus derive a mean-zero process $(a_{t})_{t\in\mathbb{N}}$ with the property that $\E[a^{2}_{t}] = \mrm{Var}[a_{t}]$.  Under the assumption of time-invariant volatility, the series $a_{t}^{2}$ should exhibit no autocorrelation at any lag $\ell\geq1$.  This assumption motivates tests for ARCH effects, that is, tests for the clustering of volatility.  These tests explore the alternative hypothesis that $\sigma_{t}^{2}$ is not only a time-varying parameter but furthermore a function of past squared residuals of the mean model.  In particular, the ARCH($m$) model is an autoregressive model in which $\sigma_{t}^{2}$ is a deterministic function of the past $m$ values of $r_{t}^{2}$.  The GARCH($m,s$) framework take this one step further by modeling $\sigma_{t}^{2}$ as a linear combination of the past $m$ values of $r_{t}^{2}$ and well as the past $s$ values of $\sigma_{t}^{2}$.  In functional form, a GARCH process (sometimes called a strong GARCH process \citep[p. 19]{francq2019garch}) is given by

\begin{align*}
&\sigma_{t}^{2} = \omega + \sum^{m}_{k=1}\alpha_{k}a^{2}_{t-k} + \sum_{j=1}^{s}\beta_{j}\sigma_{t-j}^{2}\\
&a_{t} = \sigma_{t}\epsilon_{t}\\
&\epsilon_{t} \simiid E[\epsilon_{t}]=0, Var[\epsilon_{t}] = 1\\
&\forall k,j, \alpha_{k},\beta_{j}\geq 0\\ 
&\forall t, \omega, \sigma_{t} > 0 \text { .} 
\end{align*}
Assuming further that $\sigma^{2}_{t}$ depends on a vector of exogenous covariates $\x_{t}$, we have a  GARCH-X$(m,s)$.  The volatility equation then becomes 

\begin{align}
\sigma_{t}^{2} = \omega+ \sum^{m}_{k=1}\alpha_{k}a^{2}_{t-k} + \sum_{j=1}^{s}\beta_{j}\sigma_{t-j}^{2} + \gamma^{T}\x_{t} \text{ .}\label{GARCH-X}
\end{align}


\subsection{Model setup}
\label{modelsetup}
We will suppose that a researcher has multivariate time series data $\y_{i,t} = (r_{i,t}$, $\x_{i,t}$), $t = 1,$ $\ldots,  T_i$, $i = 1, \ldots, n+1$, and  $\x_{i,t}$ is a vector of covariates such that $\x_{i,t}|\mathcal{F}_{i,t-1}$ is deterministic.  Suppose that the analyst is interested in forecasting the volatility of $r_{1,t}$, the first time series in the collection, which we will denote \textit{the time series under study}.  We require that each time series $\y_{i,t}$ is subject to an observed news event following $T^*_i \leq T_{i} + 1$ and before witnessing $T^*_i+1$.  We are implicitly leveraging the fact that financial assets are heavily traded during market hours, yet only thinly traded (if traded at all) outside market hours.  In contrast, the arrival of market-moving news does not obey any such restrictions.  In light of the foregoing, we can denote our collection of GARCH-X volatility equations of interest using the following notation

\begin{align*}
&\sigma_{i,t}^{2} = \omega_{i} + \sum^{m_{i}}_{k=1}\alpha_{i,k}a^{2}_{i,t-k} + \sum_{j=1}^{s_{i}}\beta_{i,j}\sigma_{i,t-j}^{2} + \gamma_{i}^{T} \x_{i,t} \text{ }. \\
\end{align*}
Let $I(\cdot)$ be an indicator function.  Let $T_i$ denote the time length of the time series $i$ for $i = 1, \ldots, n+1$, and let $T_i^*$ denote the largest time index prior to the arrival of the news shock, with $T_i^* < T_i$, to ensure that there is at least one post-shock realization for each series $i$.  Let $\delta, \x_{i,t} \in \mathbb{R}^{p}$.  Let $\mathcal{F}$ with a single-variable subscript denote a univariate, time-invariant $\sigma$-algebra, and $\mathcal{F}_{i,t}$ denote the canonical product filtration for donor $i$ at time $t$.  Let $D^{return}_{i,t} = I(t \in \{T_i^* + 1,...,T_i^* + L_{i, return}\})$ and $D^{vol}_{i,t} = I(t \in \{T_i^* + 1,...,T_i^* + L_{i, vol}\})$ and $L_{i,return},L_{i,vol}$ denote the lengths of the log return and volatility shocks, respectively.  For $t= 1, \ldots, T_i$ and $i = 1, \ldots, n+1$, the model $\mc{M}_1$ is defined as 
\begin{align*}
  \mc{M}_1 \colon \begin{array}{l}
     \sigma^{2}_{i,t} = \omega_{i} + \omega^{*}_{i,t} + \sum^{m_{i}}_{k=1}\alpha_{i,k}a^{2}_{i,t-k} + \sum_{j=1}^{s_{i}}\beta_{i,j}\sigma_{i,t-j}^{2} + \gamma_{i}^{T} \x_{i,t} \text{ }\\[.2cm]
     a_{i,t} = \sigma_{i,t}((1-D^{return}_{i,t})\epsilon_{i,t} + D^{return}_{i,t}\epsilon^{*}_{i})\\[.2cm]
    \omega_{i,t}^{*} = D^{vol}_{i,t}[\mu_{\omega^{*}}+\delta'\x_{i,t}+ u_{i,t}],
  \end{array}
  \end{align*}
with time-invariant error structure

  \begin{align*}
    \epsilon_{i,t} &\simiid \mc{F}_{\epsilon} \text{ with }  \; \mrm{E}_{\mc{F}_{\epsilon}}(\epsilon) = 0, \mrm{Var}_{\mc{F}_{\epsilon}}(\epsilon)  = 1  \\
    \epsilon^{*}_{i,t} &\simiid \mc{F}_{\epsilon^{*}} \text{ with }  \; \mrm{E}_{\mc{F}_{\epsilon^{*}}}(\epsilon) = \mu_{\epsilon^{*}}, \mrm{Var}_{\mc{F}_{\epsilon^{*}}}(\epsilon^{*})  = \sigma^2_{\epsilon^{*}}  \\
    u_{i,t} & \simiid  \mc{F}_{u} \text{ with }  \; \mrm{E}_{\mc{F}_{u}}(u) = 0, \mrm{Var}_{\mc{F}_{u}}(u) = \sigma^2_{u}\\
    \epsilon_{i,t} & \indep  \epsilon^{*}_{i,t}  \indep u_{i,t}
    \end{align*}
Let $\mc{M}_{0}$ denote the subclass of $\mc{M}_{1}$ models such that $\delta \equiv 0$.  Note that $\mc{M}_{0}$ assumes that $\omega^{*}_i$ have no dependence on the covariates and are i.i.d. with $\E[ \omega^{*}_i]=\mu_{\omega^{*}}$, where the lack of an index $i$ on $\mu_{\omega^{*}}$ indicates that it is shared across donors.
    
    We now introduce how a researcher can carry out predictions via distance-based weighting.  The first step is the gathering of covariates that will represent the $p$-dimensional space in which weights are generated.  The set of covariates for our models are nothing than a vector of observable covariates that parameterize a shock.  A collection of $n$ such vectors yields a matrix of $n$ columns, $\textbf{V}_{t}$.  Formally speaking, the shocks are given by the equation

    \begin{align*}
     \vec{\omega} = \vec{\mu}_{\omega^{*}} + \textbf{V}_{t}\delta + \vec{u}_{t}  
    \end{align*}
    
    where were have suppressed the $D^{vol}_{i,t}$ notation for simplicity.  Suppose that for each of the $n$ donors, we have available $p$ distinct covariates in the functional form of the shock.  The covariate set could take the form of a $p \times n$ matrix including realized volatility and implied volatility, as well the volume of traded assets on any day preceding the shock.  Covariates chosen for inclusion in a given volatility profile may be levels, log differences in levels, percentage changes in levels, or absolute values thereof, among many choices.

      Ideally, $\textbf{V}_{t}$ will display `balance' in that $p$ covariates exist for each of the $n$ donors.  In practice, missing values, corrupted values, or unacceptably extreme or noisy estimates may necessitate some sort of matrix completion, a problem that we do not tackle in this work.  We now turn to the next section, where $\textbf{V}_{t}$ is employed in a procedure to arrive at a forecast adjustment.

\section{Methodology for Similarity-based Parameter Correction}

\subsection{Forecasting}

So far we have presented a setting and a data generating process for which we intend to furnish a statistical method.  Now, we present two forecasts for the time series, the latter of which benefits from the proposed method.  In particular, we present the GARCH forecast, also called the \textit{unadjusted} forecast, as well as the \textit{adjusted} forecast, which differs by an additive term:

\begin{align*}
  \text{Forecast 1: } & \hat\sigma^{2}_{unadjusted} = \hat\E[\sigma^{2}_{1,T_{1}^{*}+1}|\mathcal{F}_{T^{*}}] && = && \hat\omega_{i} + \sum^{m_{i}}_{k=1}\hat\alpha_{i,k}a^{2}_{i,t-k} + \sum_{j=1}^{s_{i}}\hat\beta_{i,j}\sigma_{i,t-j}^{2} + \hat\gamma_{i}^{T} \x_{i,t}\\
  \text{Forecast 2: } & \hat\sigma^{2}_{adjusted} = \hat\E[\sigma^{2}_{1,T_{1}^{*}+1}|\mathcal{F}_{T^{*}}] + \hat\omega^{*} && = && \hat\omega_{i} + \sum^{m_{i}}_{k=1}\hat\alpha_{i,k}a^{2}_{i,t-k} + \sum_{j=1}^{s_{i}}\hat\beta_{i,j}\sigma_{i,t-j}^{2} + \hat\gamma_{i}^{T} \x_{i,t} + \hat\omega^{*} \text{ .}
\end{align*}

A GARCH model is an ARMA on the squares of the scalar time series $a^{2}_{t}$ \citep[][p. 18, p. 46]{tsay2005analysis,francq2019garch}, assuming that $a_{t}$ satisfies fourth-order stationarity.  This fact matters for forecasting because the $h$-step-ahead forecasting function for a GARCH model is, just like for an ARMA model, the conditional expectation function, $\mathbb{E}[ \sigma^{2}_{i,T^{*}+h} | \mathcal{F}_{T^{*}}]$, or practically speaking, the estimate thereof, $\hat{\mathbb{E}}[ \sigma^{2}_{i,T^{*}+h} |\mathcal{F}_{T^{*}}]$ \citep{zivot2009practical}.  Here we have presented one-step-ahead forecasts for a GARCH-X(1,1), without loss of generality.  For $h=2,3,4,...$, the conditional expectation is computed recursively, as is standard for iterative autoregressive forecasts.

\subsection{Excess Volatility Estimators}
    \label{Excess Volatility Estimators}
   
    The problem of aggregating estimated donor shocks begins with the data constraints.  Taking the estimated shocks as a given, we essentially observe the pair $(\{\hat\omega^{*}_{i}\}^{n+1}_{i=2},\{\textbf{v}_{i}\}^{n+1}_{i=2})$.  We wish to recover weights $\{\weight_{i}\}^{n+1}_{i=2} \in \Delta^{n}$ leading to favorable forecasting properties.  These weights are used to compute $\hat\omega^{*} \coloneq \sum^{n+1}_{i=2}\weight_{i}\hat\omega^{*}_{i}$, our forecast adjustment term.  Since the weights $\{\weight_{i}\}_{i=2}^{n+1}$ are computed using $\mathcal{F}_{T^{*}_{i}}$, the set $\{\weight_{i}\}_{i=2}^{n+1}$ is deterministic, $\textit{modulo}$ any stochastic ingredient in the numerical methods employed to approximate $\x_{1,T^{*}}$ using a convex combination of donor covariates.  We say more about the properties of the shocks $\omega^{*}_{i}$ in section $\ref{SVF_properties}$. 

    Following \citet{abadie2003economic,abadie2010synthetic}, let $\|\cdot\|_{\textbf{S}}$ denote any semi-norm on $\mathbb{R}^{p}$, and define

    \begin{align*}
    \{\pi\}_{i=2}^{n+1} = \argmin_{\pi}\|\textbf{v}_{1} - \V_{t}\pi \|_{\textbf{S}} \text{ .}
    \end{align*}
In the forecast combination literature, it is of interest whether the weights employed to aggregate forecasts strive toward and meet various optimality criteria \citep{timmermann2006forecast,wang2023forecast}.  In our endeavor, there are at least two senses of optimal weights that one might be interested in.  First, we can think of optimal weights as a set $\{\weight_{i}\}_{i=2}^{n+1}$ such that $\omega_{1} = \sum^{n+1}_{i=2}\weight_{i}\hat\omega_{i}$, i.e., $\omega_{1}$ is recovered perfectly, as it belongs to convex hull of the estimated shocks. However, $\omega_{1}$ is never revealed to the practitioner, and hence there is no way of verifying the extent to which this condition is satifised.

A more promising aim is finding weights such that $\textbf{v}_{1} = \sum^{n+1}_{i=2}\weight_{i}\textbf{v}_{i}$, meaning that the volatility profile of the time series under study lies within the convex hull of the donor volatility profile.  This condition underwrites asymptotic results in \citet{abadie2010synthetic}, and the intuition there extends to this work: if the shock is parameterized by an affine function of covariates, then finding a linear combination that recreates the shock should serve us well.  Because the method proposed uses a point in $\Delta^{n}$, it is important to head-off possible confusion.  What we are proposing is not a forecast combination method.  What we are aggregating and weighting (not combining) are subcomponents of forecasts, not forecasts themselves.  Moreover, from a broader perspective, forecast combination is an inapt term for what is being proposed here.  First, the donor time series do not provide forecasts, nor would forecasts be needed for random variables that have already been realized.  Second and more fundamentally, the theoretical underpinnings of forecast combination, while diverse \citep{wang2023forecast}, are distinct from the setting presumed in this work, where the model family is not in doubt but the parameter values and how they prevail must be learned.

Supposing that we can define optimal weights, are they necessarily unique?  \cite{lin2021minimizing} discuss sufficient conditions for uniqueness as well as the implications of non-uniqueness.  \cite{abadie2022synthetic} invoke the Carath\'eodory Theorem to argue for the sparseness of the weight vector.  We make additional comments as well. $(\mathbb{R}^{n}, \|\cdot\|)$ is a Chebyshev space, and hence for any element x and any convex set $C\subset \mathbb{R}^{n}$, there exists a unique element $y\in C$ that minimizes $\|x-y\|$.  However, the pre-image of $y$ with respect to a particular operator and constraint set might not be unique.  Let $p', n'$ denote the number of linearly independent rows of $\V_{t}$ and linearly independent columns of $\V_{t}$, respectively.  Let col($\cdot$) denote the column space of a matrix, and let Conv($\cdot$) denote the convex hull of a set of column vectors.  \\

    \begin{center}
      \begin{tabular}{ | m{3em} | m{7cm}| m{7cm} | } 
        \hline
        & $\textbf{v}_{1}\in \text{Conv}(\text{col}(\V_{t}))$ & $\textbf{v}_{1} \notin \text{Conv}(\text{col}(\V_{t}))$\\ 
        \hline
        $p' \geq n'$ & Perfect fit; fit unique & Fit not perfect; fit unique \\
        \hline
        $p' < n'$ & Perfect fit, not necessarily unique, Carath\'eodory Theorem applies& Fit not perfect, not necessarily unique, Carath\'eodory Theorem applies \\ 
        \hline
      \end{tabular}
      \end{center} 

Comparison with least-squares estimation is illustrative.  Consider least-squares for the $n$-vector of estimated volatlity shocks $\hat\omega^{*}$:

\begin{align*}
\vec{w}_{OLS}=\argmin_{{w}} \|\hat{\omega}^{*} - w^{T}\textbf{V}_{t}\|_{2}
\end{align*} 
One immediately visible problem is that this optimization problem is an optimization problem over $p$-vectors $\vec{w}$ --- i.e. over linear combinations of the covariates, whereas what we seek is an $n$-vector --- a linear combination of donors.  Additionally, there is no guarantee that $\vec{w}_{OLS}$ would perform poorly as a tool for producing $\hat\omega^{*}$, but given the small number of donors supposed in our setting, it is risky.
    \subsection{Ground Truth Estimators}
    \label{Ground Truth Estimators}
    
    The time-varying parameter $\sigma^{2}_{t}$ is a quantity for which even identifying an observable effect in the real world is far more challenging.  In this work, we use a common estimator of the variance called realized volatility (RV), one which has the virtue of being ``model-free'' in the sense that it requires no modeling assumptions \citep{andersen2010stochastic}.  The realized variance itself can be decomposed into the sum of a continous component and a jump component, with the latter being less predictable and less persistent \citep{andersen2007roughing}, cited in \citet{de2006forecasting}, two factors that further motivate the method employed herein.
    
    Suppose we examine $K$ units of of time, where each unit is divided into $m$ intervals of length $\frac{1}{m}$.  We adapt the notation of \citet{andersen2008realized}. Let $p_{t} = \log{P_{t}}$, and let $\tilde{r}(t,\frac{1}{m}) = p_{t} - p_{t-\frac{1}{m}}$.  We estimate the variance of $i$th log return series using Realized Volatility of the $K$ consecutive trading days that conclude with day $t$, denoted $RV_{i,t}^{K,m}$, using
    
    $$RV_{i,t}^{K,m} = \frac{1}{K}\sum^{Km}_{v=1}\tilde{r}^{2}(v/m,1/m),$$

    where the $K$ trading days have been chopped into $Km$ equally-sized blocks.

    Assuming that the $K$ units $\tilde{r}(t, 1) = p_{t} - p_{t-1}$ are such that $\tilde{r}(t, 1) \simiid N(\mu, \delta^{2})$, it is easily verified that 
    
    $$\E[RV^{K,m}] = \frac{\mu^{2}}{m} + \delta^{2},$$
    which is a biased but consistent estimator of the variance.  We will proceed using $m = 77$, corresponding to the 6.5-hour trading day chopped into 5-minute blocks, with the first block omitted in order to ignore unusual trading behavior at the start of the day.

\subsection{Loss Functions}\label{loss_function}

We are interested in point forecasts for $\sigma^{2}_{1,T^{*}+h}|\mathcal{F}_{T^{*}}$, $h=1,2,...,$ the $h$-step ahead conditional variance for the time series under study.  Let $L^{h}$ with the subscripted pair $\{$prediction method, ground truth estimator$\}$, denote the loss function for an $h$-step-ahead forecast using a given prediction function and ground truth estimator.  For example, if we suppress the time index, the one-step-ahead MSE using our method and Realized Volatility as the ground truth is

$$ \text{MSE}^{1}_{\text{SVF, RV}} = (\hat\sigma^{2}_{SVF} - \hat\sigma^{2}_{RV})^{2}$$
Also of interest in absolute percentage error for an $h$-step-ahead forecast, defined as

\[ 
\text{APE}^{h}_{method, ground truth} = \frac{|\hat\sigma^{2}_{h, method} - \hat\sigma^{2}_{h, ground truth}|}{\hat\sigma^{2}_{h, ground truth}}
\]
Finally, we introduce the QL (quasi-likelihood) Loss \citep{brownlees2011practical}:

\[ 
\text{QL}^{h}_{method, ground truth} = \frac{\hat\sigma^{2}_{h, ground truth}}{ \hat\sigma^{2}_{h, method}} - \log{\frac{\hat\sigma^{2}_{h, ground truth}}{ \hat\sigma^{2}_{h, method}}} -1 \text{ .}
\]
What distinguishes QL Loss is that it is multiplicative rather than additive.  This has benefits, both practical and theoretical.  As \citet{brownlees2011practical} explain, the technical properties of the QL Loss allow researchers to compare forecasts across heterogeneous time series, whereas additive loss functions like MSE unfairly penalize forecasts made under market turbulence.  For this reason and others, we proceed to evaluate the method, both in simulations and real data examples, using the QL loss.

\section{Properties of Volatility Shocks and Shock Estimators}\label{SVF_properties}

The model $\mc{M}_1$ is defined by a volatility equation and mean equation, as is any GARCH model.  The decision to model the volatility shock $\omega^{*}_{i}$ as an additive random effect is straightforward.  However, the decision to model the level effect $\epsilon^{*}_{i,t}$ as a temporary rupture in the otherwise i.i.d. sequence of innovations $\epsilon_{i,t}$ stands in need of deeper justification.  One way of arguing for this choice is that, in a discrete time series model, if we assume the arrival of news in the time between $T^{*}$ and $T^{*}+1$, we do not have an easy way to express a conditional distribution of the innovation $\epsilon_{T^{*}+1}$ given the overnight arrival of information.  Using $\epsilon^{*}_{i,t}$ thus breaks this impasse.  This justification also explains why we do not parameterize the level shock at $T^{*}+1$ as a sum of two shocks, $\epsilon_{i,T^{*}+1}$ and $\epsilon^{*}_{i,T^{*}+1}$, which would represent the level shock as generated by two independent sources of stochasticity.  To do so would be inelegant and would also lack motivation as a practical level.  While we want to model the shock at $T^{*}+1$ as potentially large in absolute value, we also want to retain the property of a unitary source of noise.

Note that under the popular GARCH(1,1), a dual level-volatility shock has an marginal effect on the conditional variance $\sigma^{2}_{i,t}$ that reflects the geometric decay of innovations in autoregressive models.  As usual, assume $\alpha+\beta < 1$.  Furthermore, assume that both the volatility shock and the level shock are of length one only, and consider a circumstance with no exogenous covariate $\x_{i,t}$. Assume also that $r\geq 2$, which is necessary in order to isolate the effects of the level shock $\epsilon^{*}_{i,t}$.  Then

\begin{align}
\sigma^{2}_{i,T^{*}+r+1} | \mathcal{F}_{T^{*}+r} & = \omega_{i} + \alpha_{i} a_{T^{*}+r}^{2} + \beta_{i}\sigma^{2}_{i,T^{*}+r} \label{eq0}\\
& = \omega_{i} + \alpha_{i}(\sigma_{i,T^{*}+r}\epsilon_{T^{*}+r})^{2} + \beta_{i}\sigma^{2}_{i,T^{*}+r}\notag \\
& = \omega_{i} + \sigma^{2}_{i,T^{*}+r}(\alpha_{i} (\epsilon_{T^{*}+r})^{2} + \beta_{i}) \text{ .}\notag 
\end{align}

In Equation \eqref{eq0}, observe that $\omega_{i,t}^{*}$ and $\epsilon^{*}_{i,t}$ each appear at most once, through the term $\sigma^{2}_{T^{*}+r}$.  This might lead one to suspect  geometric decay of the shocks $\omega_{i,t}^{*}$ and $\epsilon^{*}_{i}$.  Such a suspicion is easier to substantiate by examining the conditional expectation of the variance, $\mathbb{E}[ \sigma^{2}_{i,T^{*}+r+1} |\mathcal{F}_{T^{*}+r}]$, which also happens to be the principal forecasting tool for a GARCH model \citep{zivot2009practical}.  Indeed, if we assume unit variance for all $\epsilon_{i,t}$ except, of course, $\epsilon^{*}_{i,t}$, then we have

\begin{align*}
\mathbb{E}[ \sigma^{2}_{i,T^{*}+r+1} |\mathcal{F}_{T^{*}+r}] & = \mathbb{E}[\omega_{i} + \alpha a_{T^{*}+r}^{2} + \beta\sigma^{2}_{i,T^{*}+r} |\mathcal{F}_{T^{*}+r}] \\
& = \omega_{i} + \mathbb{E}[\alpha(\sigma_{i,T^{*}+r}\epsilon_{T^{*}+r})^{2} |\mathcal{F}_{T^{*}+r}] + \beta\sigma^{2}_{i,T^{*}+r} \\
& = \omega_{i} + \alpha\sigma_{i,T^{*}+r}^{2} + \beta\sigma^{2}_{i,T^{*}+r} \tag{Due to the unit variance assumption}\\
& = \omega_{i} + \sigma^{2}_{i,T^{*}+r}(\alpha + \beta) \text{ .} \\
\end{align*}

By repeated substitution, in conditional expectation, the shock is $\mathcal{O}((\alpha+\beta)^{r})$.  We generalize this observation in the following proposition.

\begin{prop}
Let $a_{t}$ be a mean-zero time series obeying a GARCH(1,1) specification with unit-variance errors, all prior to the arrival of a volatility shock of length $L_{\text{vol}} \geq 1$ and level shock of length $L_{\text{return}}\geq 1$ at some time $T^{*}+1$.  Then for any $r$ such that $r \geq \text{max}\{L_{i, vol},L_{i, level}\} + 1$, 

\begin{align*}
\mathbb{E}[ \sigma^{2}_{i,T^{*}+r+1} |\mathcal{F}_{T^{*}+r}] & = \omega_{i} + (\alpha + \beta)\sigma^{2}_{i,T^{*}+r}
\end{align*}
\end{prop}

\begin{proof-of-proposition}
We claim

\begin{align}
\mathbb{E}[ \sigma^{2}_{i,T^{*}+r+1} |\mathcal{F}_{T^{*}+r}] & = \mathbb{E}[\omega_{i} + \alpha a_{T^{*}+r}^{2} + \beta\sigma^{2}_{i,T^{*}+r} |\mathcal{F}_{T^{*}+r}] \label{eq1}\\
& = \omega_{i} + \mathbb{E}[\alpha(\sigma_{i,T^{*}+r}\epsilon_{T^{*}+r})^{2} |\mathcal{F}_{T^{*}+r}] + \beta\sigma^{2}_{i,T^{*}+r} \label{eq2}\\
& = \omega_{i} + \alpha\sigma_{i,T^{*}+r}^{2} + \beta\sigma^{2}_{i,T^{*}+r} \label{eq3}\\
& = \omega_{i} + (\alpha + \beta) \sigma^{2}_{i,T^{*}+r}\label{eq4} \text{ .}
\end{align}
The volatility equation of a GARCH(1,1) dictates that for any $r$, the one-step-ahead volatility is given by the expression inside the expectation in \eqref{eq1}.  By the mean-model assumption of a GARCH(1,1), we have $a_{i,t} = \sigma_{i,t}\epsilon_{i,t}$, and hence by substituting $\sigma_{i,t}\epsilon_{i,t}$ for $a_{i,t}$, we arrive at equation \eqref{eq2} above.  Using the unit-variance assumption on $\epsilon_{T^{*}+r}$, we can compute explicitly the expectation, yielding (\ref{eq3}).  Finally, by rearranging terms, we arrive at equation \eqref{eq4}.
\end{proof-of-proposition}
In other words, for a GARCH(1,1), once two time points removed from the longest shock length, the volatility shock and level shock can be subsumed into one.  However, prior to being two time points removed, there is no such guarantee.  For example, one can take $r = 1$ and level shock of length at least 1 to see that 

\begin{align*}
\mathbb{E}[ \sigma^{2}_{i,T^{*}+2} |\mathcal{F}_{T^{*}+1}] & = \mathbb{E}[\omega_{i} + \alpha a_{T^{*}+1}^{2} + \beta\sigma^{2}_{i,T^{*}+1} |\mathcal{F}_{T^{*}+1}] \\
& = \omega_{i} + \mathbb{E}[\alpha(\sigma_{i,T^{*}+r}\epsilon^{*}_{T^{*}+1})^{2} |\mathcal{F}_{T^{*}+1}] + \beta\sigma^{2}_{i,T^{*}+1} \\
& = \omega_{i} + \alpha\sigma^{2}_{i,T^{*}+1}(\mu^{2}_{\epsilon^{*}} + \sigma^{2}_{\epsilon^{*}}) + \beta\sigma^{2}_{i,T^{*}+1} \\
& = \omega_{i} + \sigma^{2}_{i,T^{*}+1}(\alpha(\mu^{2}_{\epsilon^{*}} + \sigma^{2}_{\epsilon^{*}}) + \beta)\text{ .}
\end{align*}
where $(\alpha(\mu^{2}_{\epsilon^{*}} + \sigma^{2}_{\epsilon^{*}}) + \beta)$ may be greater than 1, permitting explosive behavior, at least in the short term.  After both shocks have been exhausted, their influence disappears quickly.  This short-memory effect has implications for the method being developed herein.  First, there may be different risks associated with over/underestimating level shock and vol shock lengths.  Estimation of effects in donor pool should err on the side of underestimating, not overestimating, the length of the max shock, since overestimation of the shock length brings with it the risk of underestimating $\omega^{*}$.  Second, a practitioner of the method needs some idea of how long the operator expects the respective shocks in the time series under study to last.  There are couple of obvious strategies: take all the donors, and over all the donor shock lengths, take the minimium.  Alternatively, one could take the maximum.

\subsection{Consistency of the fixed effect estimators}

\begin{prop}\label{omega_consistency}
Assume
\begin{enumerate}
  \item For each $i$, $\{a_{i,t}\}_{t=0,...,T_i}$ obeys a GARCH-X($m,s$), as laid out in Equation $\eqref{GARCH-X}$, with volatility shocks found in $\mc{M}_{1}$, where $T_i$ is the length of the $i$th series.
  \item For each $i, \{\omega_{i,t}^{*}\}_{t=0,...,T_i}$ is potentially non-zero at $\{T^{*}_{i}+1,... ,T^{*}_{i}+k\}$, $\omega_{i,T^{*}+1}^{*}\equiv...\equiv\omega_{i,T^{*}+k}^{*}$, and zero otherwise, where the arrival of $T_{i}^{*}$ is governed by a time-invariant distribution on $\{a_{i,t}\}_{t=0,...,T_i-1}$. \label{stationarity_of_omega_i_t}
  \item The conditions in Assumption 0 of \citet{han2014asymptotic} prevail.
\end{enumerate}
Then for any $i$, $\hat\omega_{i,t}^{*} \xlongrightarrow{p} \omega_{i,T^{*}+1}^{*}$ as $t\rightarrow\infty$.
\end{prop}

\begin{lem}
  Under assumption \ref{stationarity_of_omega_i_t}, for each $i, i=1,...,n+1$,  $\{\omega_{i,t}^{*}\}_{t=0,...,T_i}$ is a strictly stationary series.
\end{lem}

\begin{proof-of-lemma}
Since the shock is assumed to arrive uniformly at random for each $i$, $1 \leq i \leq n + 1$, and last for a discrete number of indices, the sequence $\{\omega_{i,t}^{*}\}_{t=0,...,T_i}$ is governed by a distribution $F_{\{\omega_{i,t}^{*}\}_{t=0,...,T_i}}$ that is invariant to shifts in time.
\end{proof-of-lemma}

\begin{proof-of-proposition}
The result follows from the consistency proof of the QMLE in GARCH-X models, as established by \citet{han2014asymptotic}. 
\end{proof-of-proposition}

  \subsection{Consistency of the Forecast Function}
  \begin{prop}\label{sigma_consistency}
    Assume
    \begin{enumerate}
      \item All conditions listed in Proposition \ref{omega_consistency}.
      \item There exist weights $\{\pi_{i}\}_{i=2}^{n+1}$ such that $\textbf{v}_{1,T^{*}} = \sum^{n+1}_{i=2}\weight_{i} \textbf{v}_{i,T^{*}}$.
     \end{enumerate}
  Then $\hat\sigma^{2}_{adjusted}\xlongrightarrow{p}\sigma^{2}_{1,T^{*}+1}$ as $t\rightarrow\infty$. 
  \end{prop}

\begin{proof-of-proposition}
Recall the conditional expectation of the variance for the GARCH-X($m$,$s$) model:

\begin{align*}
\E_{T^{*}}[\sigma^{2}_{i,t+1}|\mathcal{F}_{t}] = \omega_{i} + \omega^{*}_i + \sum^{m_{i}}_{k=1}\alpha_{i,k}a^{2}_{i,t-k} + \sum_{j=1}^{s_{i}}\beta_{i,j}\sigma_{i,t-j}^{2} + \gamma_{i}^{T} \x_{i,t} \text{ .}
\end{align*}

By replacing parameters with their estimates, we arrive at the prediction 

\begin{align*}
\hat\sigma^{2}_{i,t+1}|\mathcal{F}_{t} = \hat\omega_{i} + \hat\omega^{*}_i + \sum^{m_{i}}_{k=1}\hat\alpha_{i,k}a^{2}_{i,t-k} + \sum_{j=1}^{s_{i}}\hat\beta_{i,j}\hat\sigma_{i,t-j}^{2} + \hat\gamma_{i}^{T} \x_{i,t} \text { ,}
\end{align*}

which converges in probability to 

\begin{align*}
  \hat\sigma^{2}_{i,t+1}|\mathcal{F}_{t} = \omega_{i} + \omega^{*}_i + \sum^{m_{i}}_{k=1}\alpha_{i,k}a^{2}_{i,t-k} + \sum_{j=1}^{s_{i}}\beta_{i,j}\sigma_{i,t-j}^{2} + \gamma_{i}^{T} \x_{i,t}
  \end{align*}

as $t\rightarrow\infty$ by a simple application of Slutsky's Theorem.

\end{proof-of-proposition}

\subsection{Asymptotic Loss}

We now evaluate the loss and risk of our method under two scenarios: first, under arbitrary distribution of $\sigma^{2}_{t+1}$, and then second, under the assumption that the data-generating process is correctly specified.

For 1-step-ahead forecast of $\sigma^{2}_{t+1}$ where $t=T^{*}$, consider the difference 

\begin{align}
  & QL(\hat\sigma_{t+1, unadjusted}^{2},\sigma^{2}_{t+1})-QL(\hat\sigma^{2}_{t+1, adjusted},\sigma^{2}_{t+1})\notag \\
   & =(\frac{\sigma_{t+1}^{2}}{\hat\sigma^{2}_{t+1,unadjusted}} - \log{\frac{\sigma_{t+1}^{2}}{\hat\sigma^{2}_{t+1,unadjusted}} } - 1) - (\frac{\sigma_{t+1}^{2}}{\hat\sigma^{2}_{t+1,adjusted}} - \log{\frac{\sigma_{t+1}^{2}}{\hat\sigma^{2}_{t+1,adjusted}}} - 1)\notag\\
   & = \frac{\sigma_{t+1}^{2}}{\hat\sigma^{2}_{t+1,unadjusted}} - \frac{\sigma_{t+1}^{2}}{\hat\sigma^{2}_{t+1,adjusted}}+ \log{\frac{\hat\sigma^{2}_{t+1,unadjusted}}{\hat\sigma^{2}_{t+1,adjusted}}}\notag\\
   & = \frac{\sigma_{t+1}^{2}(\hat\sigma^{2}_{t+1,adjusted}-\hat\sigma^{2}_{t+1,unadjusted})}{\hat\sigma^{2}_{t+1,adjusted}\hat\sigma^{2}_{t+1,unadjusted}} + \log{\frac{\hat\sigma^{2}_{t+1,unadjusted}}{\hat\sigma^{2}_{t+1,adjusted}}}\label{QL Loss toy example} \text{ .}
\end{align}
For simplicity, we work with a GARCH(1,1) that experiences a volatility shock at a single time point for which we would like to provide a point forecast.  Then (\ref{QL Loss toy example}) can be expressed as

\begin{align*}
   &\frac{\sigma^{2}_{t+1}\hat\omega^{*}_{t+1} }{\hat\sigma^{2}_{t+1,adjusted}\hat\sigma^{2}_{t+1,unadjusted}} + \log{\frac{\hat\omega + \hat\alpha a_{t}^{2} + \hat\beta\sigma_{t}^{2}}{\hat\omega + \hat\alpha a_{t}^{2} + \hat\beta\sigma_{t}^{2} + \hat\omega^{*}_{t+1}}}\\
\end{align*}\label{QL Loss Consistency - GARCH(1,1)}
It is easily verified that as $\hat\omega^{*}_{t+1} \rightarrow 0^{+}$, the difference in the losses goes to zero.  On the other hand, as $\hat\omega^{*}_{t+1}$ becomes large, the difference in the losses turns negative, with the lesson being that $\hat\omega^{*}_{t+1}$ must be in appropriate proportion to $\sigma^{2}_{t+1}$ in order for the adjusted forecast to outperform the unadjusted forecast.  This explains why it is so important to avoid using a naive adjustment estimator, $\overline{\omega^{*}}$, the arithmetic mean of the estimated shocks.  We conclude this section with a broader result. 

\begin{prop}\label{asymptotic_consistency}
Assume the conditions in Propositions $\ref{omega_consistency}$ and $\ref{sigma_consistency}$.  Let $1\leq i\leq n+1$.  Then as $t\rightarrow \infty$
  
$$QL(\hat\sigma_{t+1, unadjusted}^{2},\sigma^{2}_{t+1})-QL(\hat\sigma^{2}_{t+1, adjusted},\sigma^{2}_{t+1})\xlongrightarrow{p} \frac{\omega_{i,t+1}^{*}}{\sigma^{2}_{t+1}-\omega_{i,t+1}^{*}} + \log{\frac{\sigma_{t+1}^{2}-\omega_{i,t+1}^{*}}{\sigma_{t+1}^{2}} } \geq 0.$$\\

Hence, a correctly specified $\mc{M}_1$ model will outperform the unadjusted forecast asymptotically.
\end{prop}

\begin{proof-of-proposition}
  First, note that the function $g:(-\infty,\omega_{i}^{*})\rightarrow \mathbb{R}$ given by $g(x) = \frac{x}{\sigma^{2}_{t+1}-x} + \log{\frac{\sigma_{t+1}^{2}-x}{\sigma_{t+1}^{2}} }$ is nonnegative, convex, obtains a minimum at $x = 0$, and being continuous, preserves consistency. The conclusion follows from fact that the model is correctly specified and consistency of $sigma^{2}_{adjusted}$ is guaranteed Proposition ($\ref{sigma_consistency}$). 
  
\end{proof-of-proposition}

\section{Numerical Examples}

In this section, we demonstrate the effectiveness of the proposed method using Monte Carlo simulations.  All simulations will use $\mc{M}_1$ volatility models and $\mc{M}_0$ models on the returns, i.e. $  D^{return}_{i,t} \equiv 0$ for each $i$ and each $t$.  We first explain the parameters to be varied, the parameters that remain fixed, and the behavior we expect to observe.  For our purposes, a parameter is defined to be fixed whenever it does not vary across any of the simulations performed, and conversely, a parameter is varying whenever it varies across at least two simulations performed.

The overarching story told by the simulations is that our method performs well as the magnitude of the signal in the shock grows in comparison to the idiosyncratic noise term of the shock.  For brevity, we refer to this as the signal-to-noise phenomenon.  However, this phenomenon is complicated somewhat by at least two nuances: additional parameters must be in place and properly sized in order to enable to signal-to-noise phenomenon, and second, certain hurdles must be met in the size of those additional parameters before the signal-to-noise story can be seen.  In short, interaction effects abound, with non-linearities as well.

  \subsection{Fixed Parameters}
Each simulation, $i=1,...n+1$ is a GARCH(1,1) process with length $T_{i}$ chosen randomly from the set of integers $\{756,...,2520\}$, corresponding to approximately 3-10 years of daily financial data.  We fixed the intercept of the processes at the $\textit{garchx}$ package default of $\omega = .2$ \citep{RePEc:pra:mprapa:100301}.  We also use the values $\alpha=.1, \beta = .82$, corresponding to a GARCH process with greater influence from past values of $\sigma^{2}_{i,t}$.
  \subsection{Varying Parameters}

  We vary exactly five parameters: $\mu_{V}, \sigma_{V}, \mu_{\delta}, \mu_{\omega^{*}}, \sigma_{u}$, each of which is critical to evaluating the method's responsiveness to changing conditions in the shock distribution. $\mu_{V}, \sigma_{V}$ govern the elements of the covariates. $\mu_{\delta}$ interacts with $\textbf{V}_{t}$ via the dot-product operation, of course.  $\mu_{\omega^{*}}$ is a location parameter for the volatility shock, and $\sigma_{u}$ is the idiosyncratic noise of the volatility shock.
  
  We add an important note about the parameter $\mu_{\delta}$, which governs a vector $\delta$ of length $p$ with elements that increase monotonically in proportion to $1,...,p$, after which they are scaled by the factor $\frac{2\cdot\mu_{\delta}}{p(p+1)}$, so that a randomly selected entry of the vector has mean $\mu_{\delta}$.  The heterogeneity of the entries is critical to simulating the plausible heterogeneity that will exist in the random effects structure.  Absent a heterogeneous vector $\delta$, the shock would not vary systematically with regard to each variable chosen for the volatility profile, which would fail to approximate the realities of financial markets.

\subsection{Evaluative Framework and Monte Carlo Results}
\begin{figure}[!h]
      \centering
      \textbf{Signal and Noise}\par\medskip
    \begin{subfigure}{.44\linewidth} 
      \centering
        \includegraphics[scale = .42]{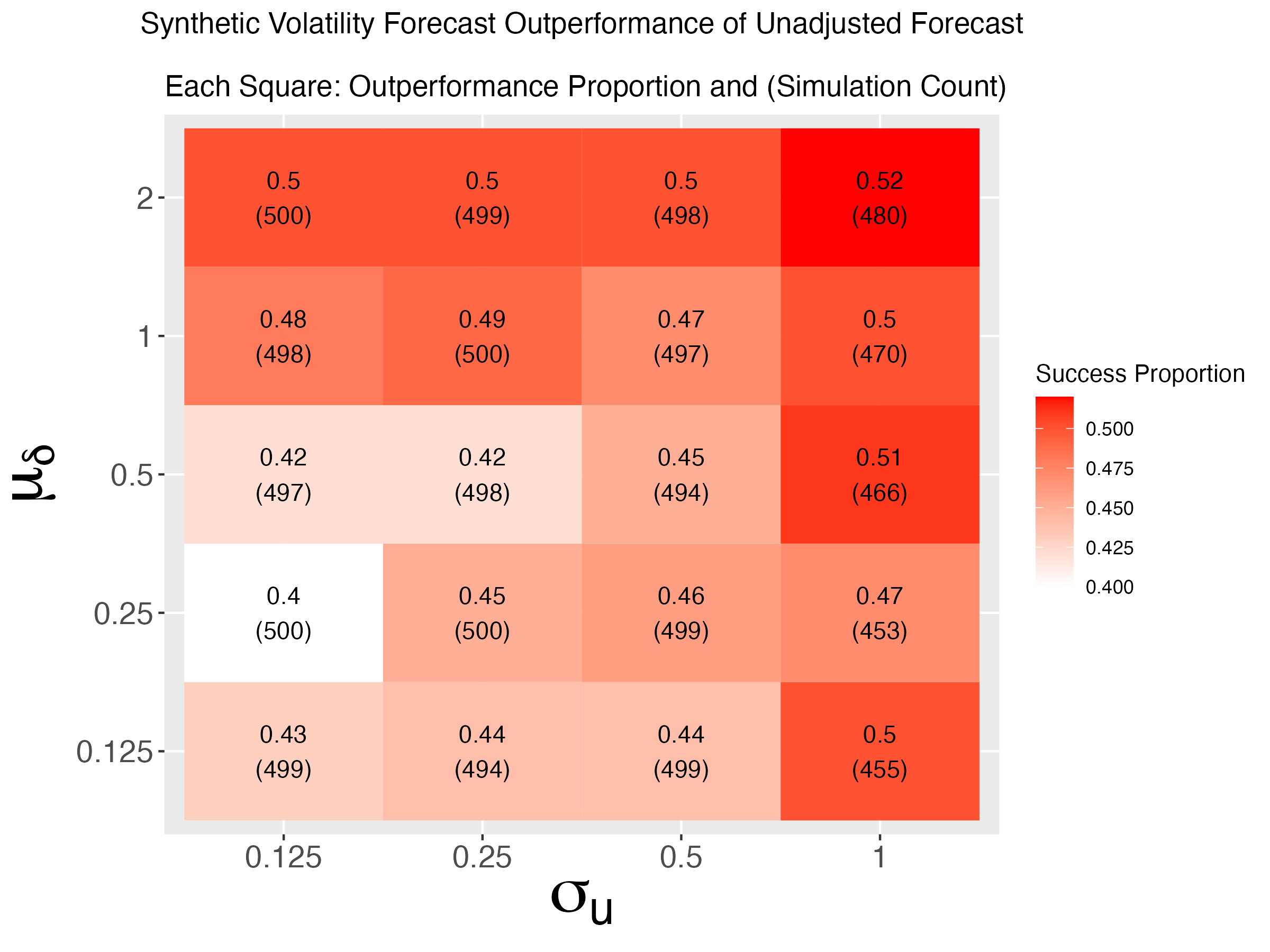}
        \caption{Fixed values: $\mu_{v} = .125, \sigma_{v} = .125, \mu_{\omega^{*}} = .125$}\label{fig:sim_1}
    \end{subfigure}\hspace{12mm} %
    \begin{subfigure}{.44\linewidth} 
      \centering
        \includegraphics[scale=.42]{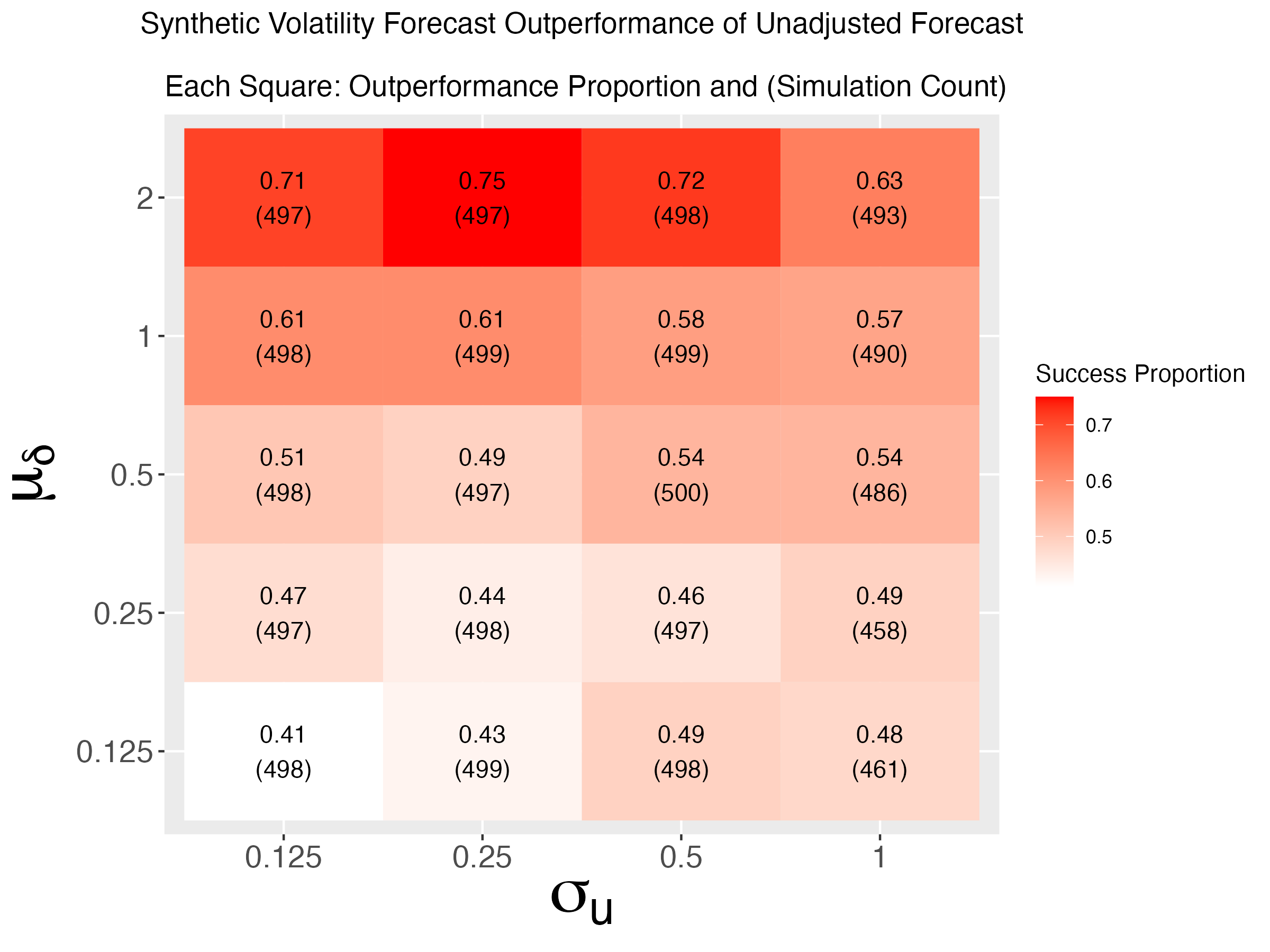}
        \caption{Fixed values: $\mu_{v} = .5, \sigma_{v} = .125, \mu_{\omega^{*}} = .125$}\label{fig:sim_2}
    \end{subfigure}

    \begin{subfigure}{.44\linewidth} 
      \centering
        \includegraphics[scale=.42]{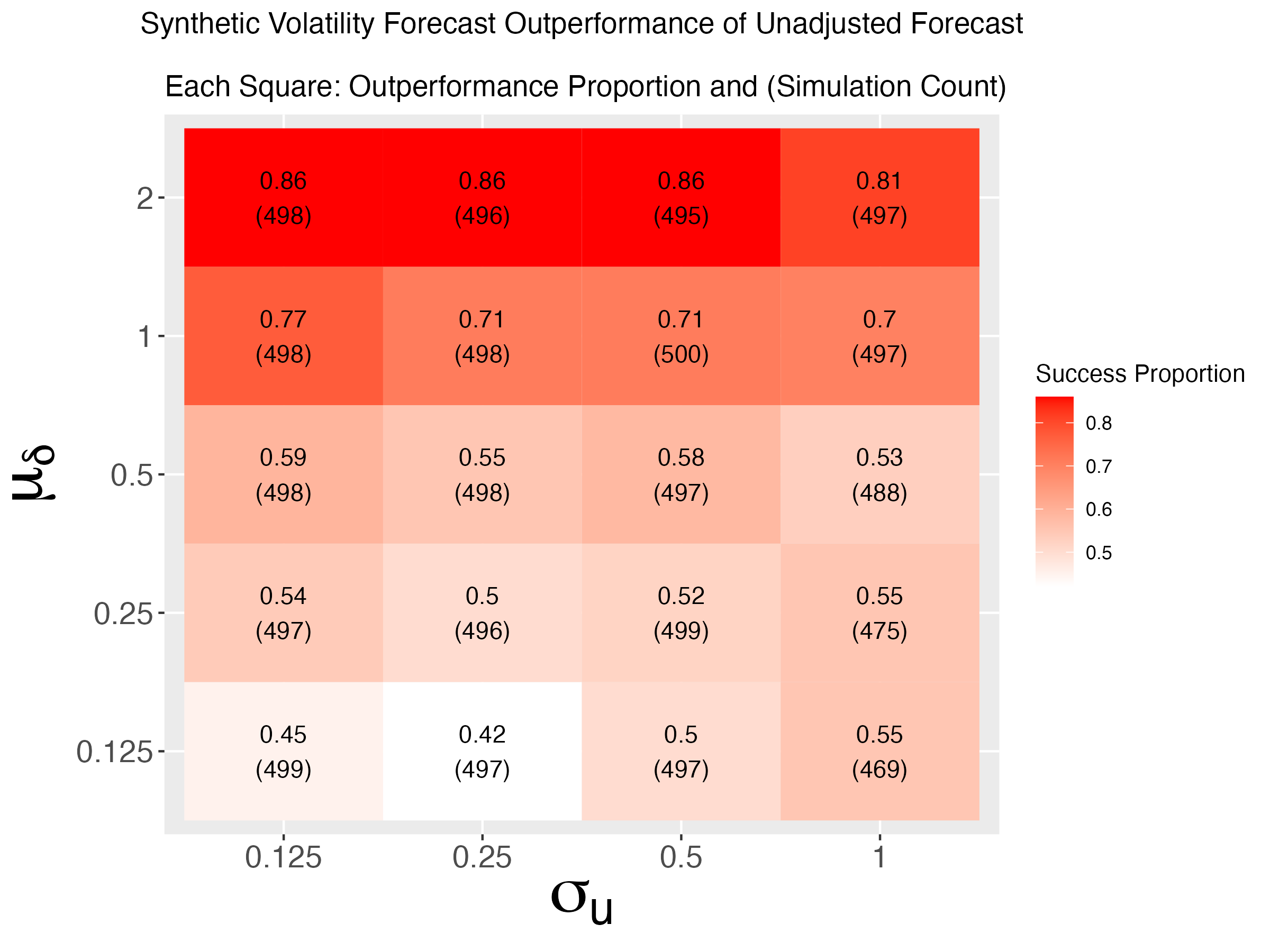}
        \caption{Fixed values: $\mu_{v} = 1, \sigma_{v} = .125, \mu_{\omega^{*}} = .125$}\label{fig:sim_3}
    \end{subfigure}
    
        \caption{In the progression from \ref{fig:sim_1} to \ref{fig:sim_2} to \ref{fig:sim_3}, we see that increasing $\mu_{\delta}$ leads to improved performance of the adjusted forecast, and this effect is intensified as $\mu_{v}$ increases.  However, it is not consistently true that an increasing noise undermines the performance of the adjusted forecast.  Rather, that phenomenon requires both large quantites of $\mu_{\delta}$ and $\mu_{v}$.}
        \label{fig:signoise}
      \end{figure}

\begin{figure}[!h]
    \centering
    \textbf{Interaction between Signal and Volatility Profile Means}\par\medskip
  \begin{subfigure}{.44\linewidth} 
    \centering
      \includegraphics[scale = .42]{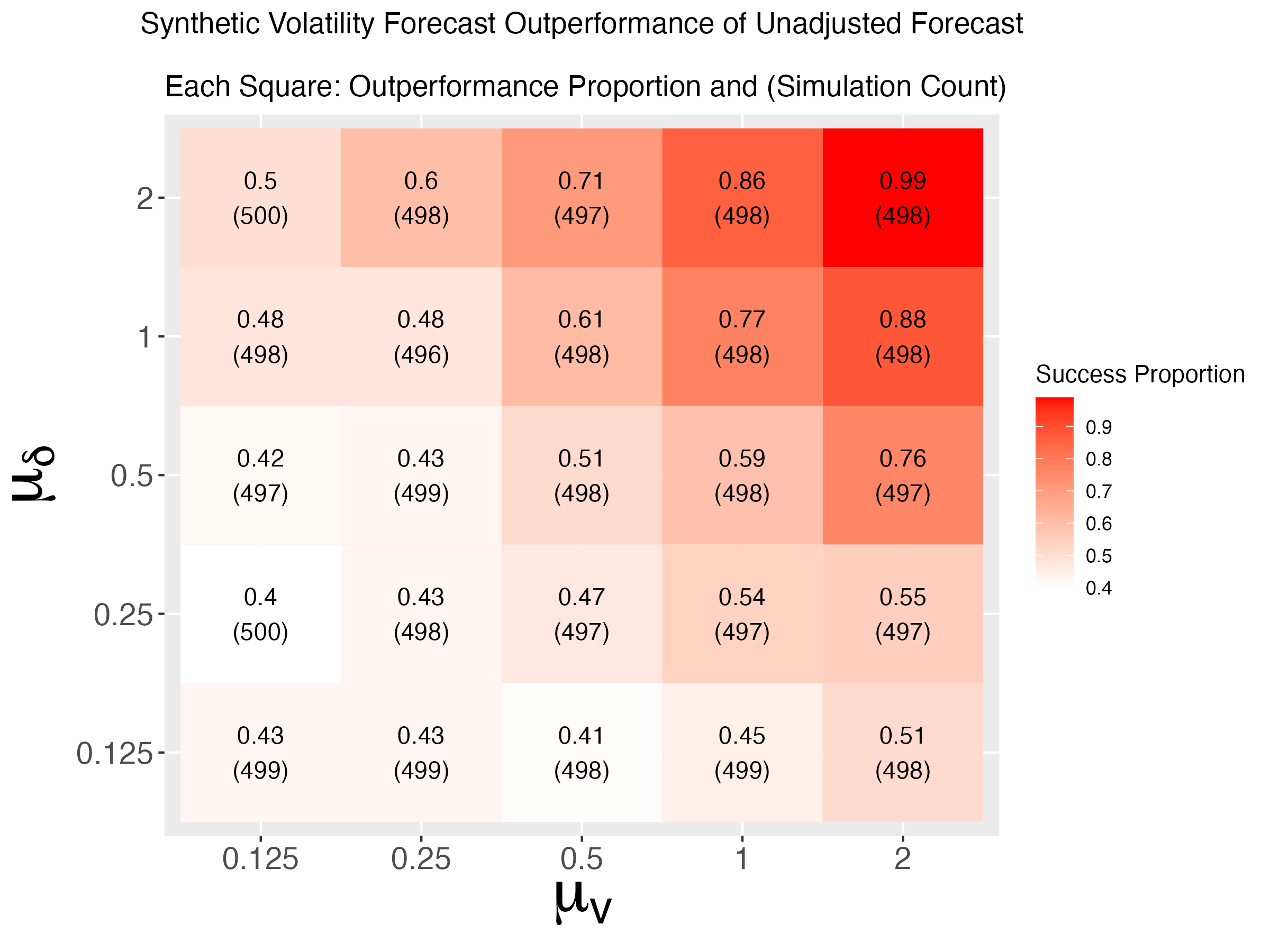}
      \caption{Fixed values: $\sigma_{v} = .125, \mu_{\omega^{*}} = .125, \sigma_{u} = .125$}\label{fig:sim_4}
  \end{subfigure}\hspace{12mm} %
  \begin{subfigure}{.44\linewidth} 
    \centering
      \includegraphics[scale=.42]{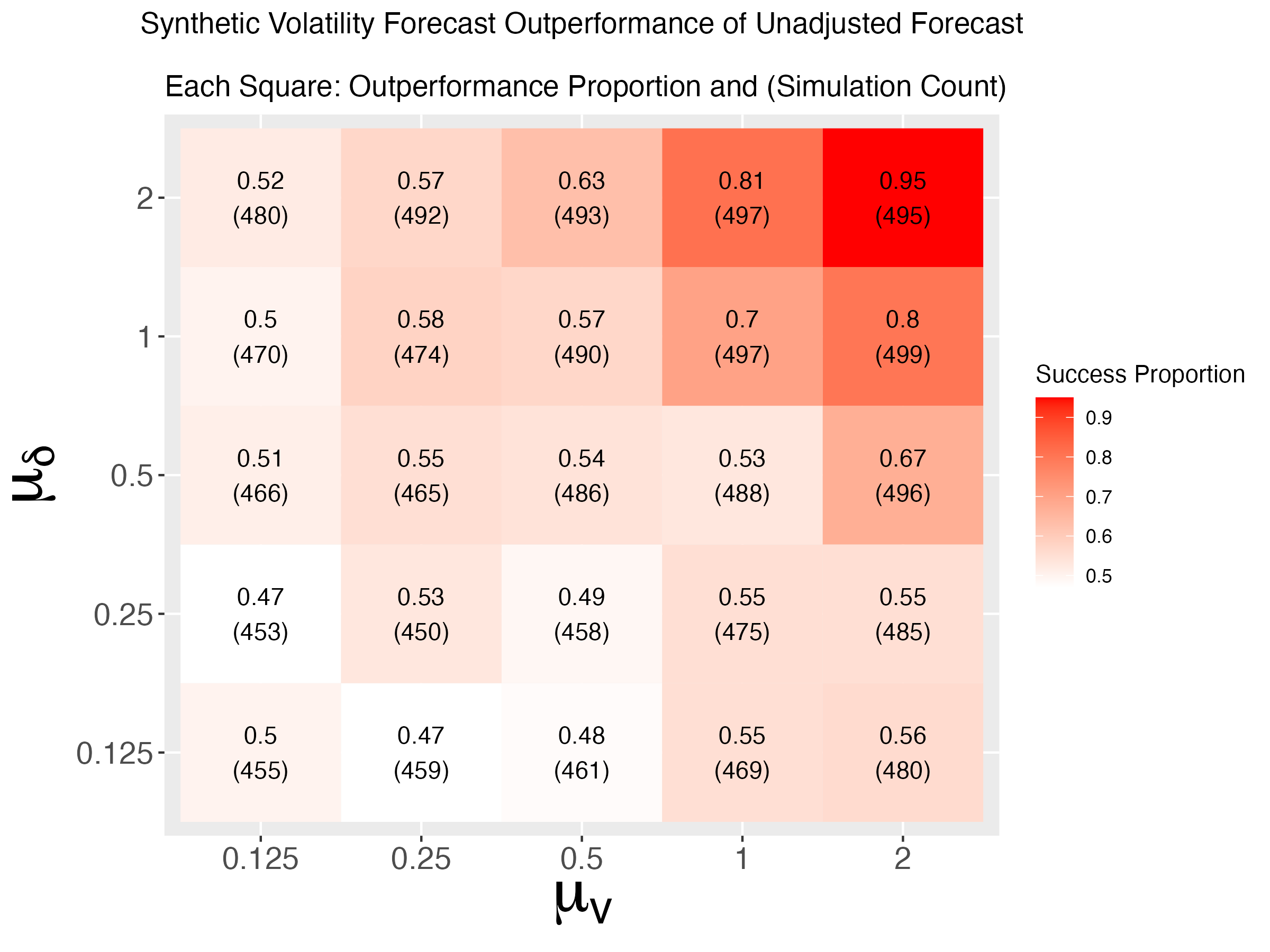}
      \caption{Fixed values: $\sigma_{v} = .125, \mu_{\omega^{*}} = .125, \sigma_{u} = 1$}\label{fig:sim_5}
  \end{subfigure}
  
      \caption{We compare the interaction of $\mu_{\delta}$ and $\mu_{v}$ at two different levels of shock noise.  In the low-noise regime, the peak performance of the adjusted forecast is higher, and the ascent is faster along both dimensions.   However, in the high-noise regime, the adjusted forecast performs well even at low levels of $\mu_{\delta}$ and $\mu_{v}$.}
      \label{fig:sig_volprof}
    \end{figure}

\begin{figure}[!h]
  \centering
  \textbf{Interaction between Shock Intercept and Shock Noise}\par\medskip
\begin{subfigure}{.44\linewidth} 
  \centering
    \includegraphics[scale = .42]{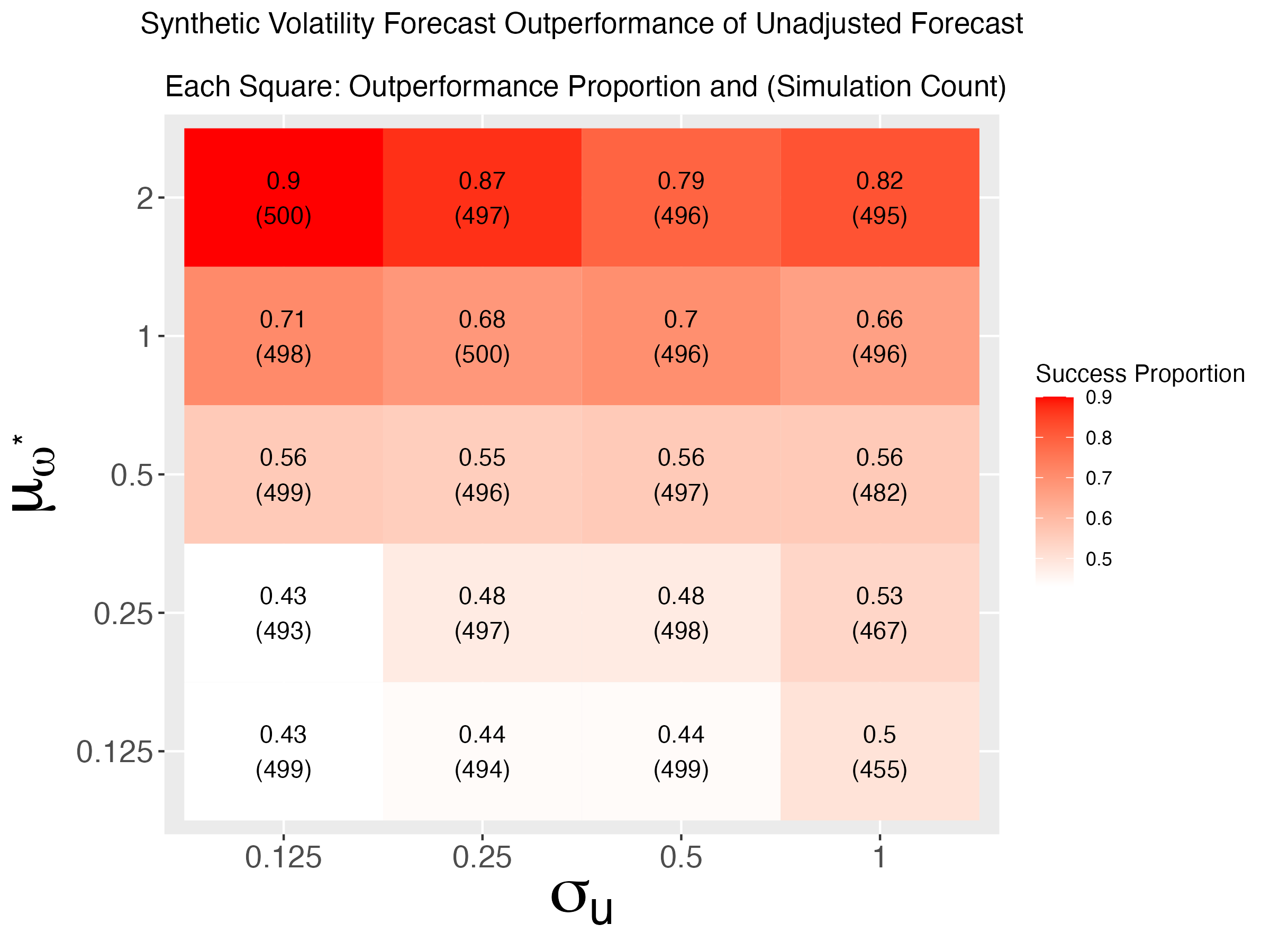}
    \caption{Fixed values: $\mu_{v} = .125, \sigma_{v} = .125, \delta = .125$}\label{fig:sim_6}
\end{subfigure}\hspace{12mm} %
\begin{subfigure}{.44\linewidth} 
  \centering
    \includegraphics[scale=.42]{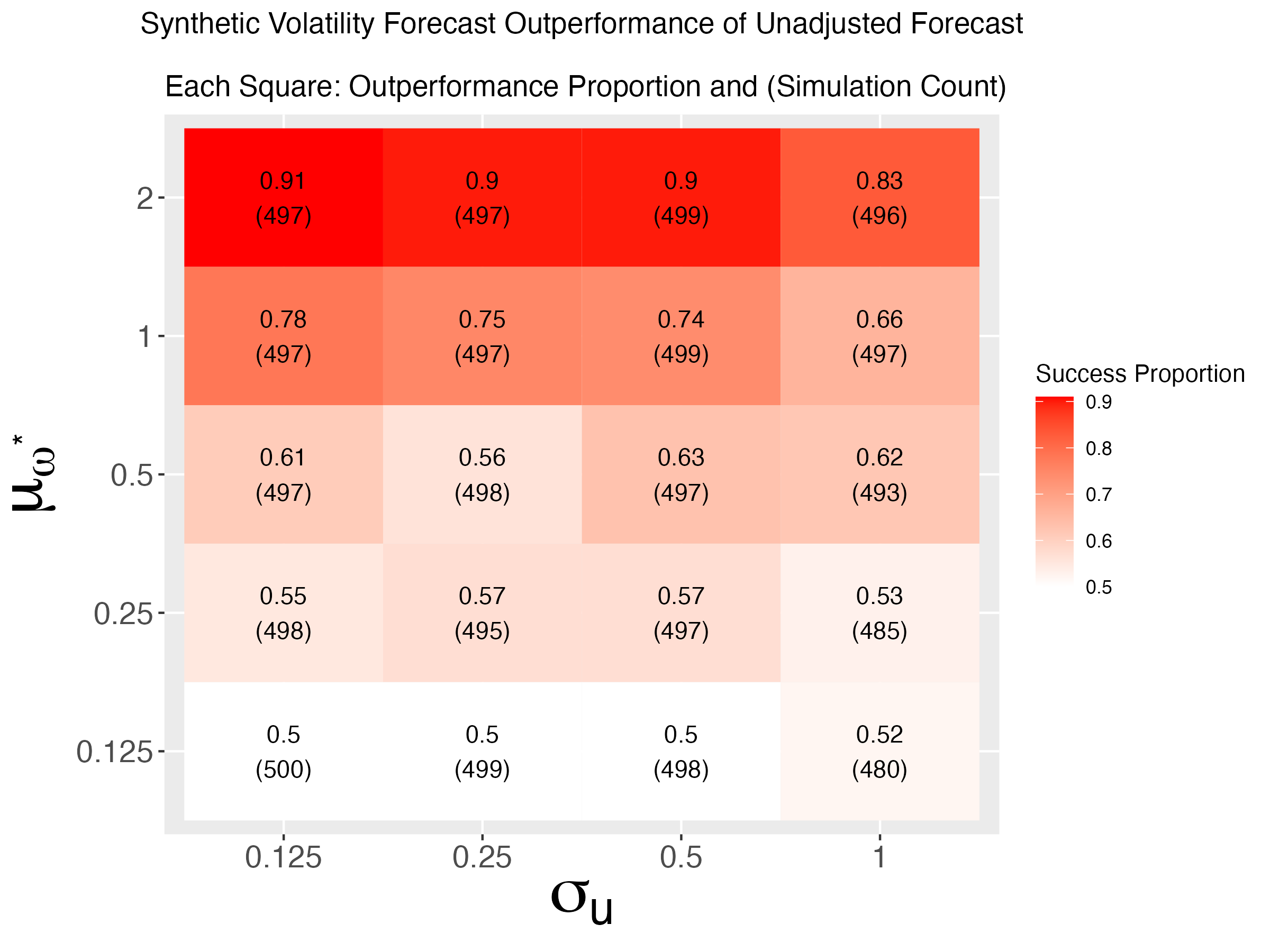}
    \caption{Fixed values: $\mu_{v} = .125, \sigma_{v} = .125, \delta = 2$}\label{fig:sim_7}
\end{subfigure}

    \caption{The interaction between the intercept $\mu_{\omega^{*}}$ and $\sigma_{u}$ suggests that the parameters behave as expected.  However, a larger value of $\delta$ in \ref{fig:sim_7} attenuates the effect of increasing noise.}
    \label{fig:intercept_noise}
  \end{figure}

Consistent with our evaluative framework declared in \ref{loss_function}, we compare adjusted and unadjusted forecasts using QL Loss, calculating the fraction of the simulations in which the adjusted forecast yields a QL Loss no smaller than that of the unadjusted forecast.  The simulations fall into three broad categories, which we will denote \textbf{Signal and Noise} (displayed in Figure \ref{fig:signoise}), \textbf{Interaction between Signal and Volatility Profile Means} (displayed in Figure \ref{fig:sig_volprof}), and lastly, \textbf{Interaction between Shock Intercept and Shock Noise} (displayed in Figure \ref{fig:intercept_noise}).  

The first category plots two parameters of the shock equation, $\mu_{\delta}$ and $\sigma_{u}$, in grids while increasing the mean of the covariate vectors $\x$ over three plots.  Recall that $\delta$ is a parameter shared between donors, whereas the vectors $\v_{i}$ in the volatility profile are i.i.d. multivariate normal.  In Figure \ref{fig:sim_1}, we see very little variation across the grid.  However, in Figures \ref{fig:sim_2} and \ref{fig:sim_3}, where $\mu_{v}$ is increased, we observe an expected phenomenon: for any fixed column of the plots, the performance of the adjusted forecast improves with increasing $\mu_{\delta}$.  A somewhat subtler phenomenon is also observed: for a fixed row of the grids in Figures \ref{fig:sim_2} and \ref{fig:sim_3}, for $\mu_{\delta} \geq .5$, increasing $\sigma_{u}$ leads to a decline in performance.  However, this is not observed for smaller levels of $\mu_{\delta}$, where the signal is too weak.

The second category of simulations examines $\mu_{\delta}$ and $\mu_{v}$, while increasing the noise $\sigma_{u}$.  The row-wise and column-wise increases in performance that we witnessed in the first category largely hold in the second category of simulations, with one similar disclaimer: the relationship is strained under large levels of noise, like $\sigma_{u} = 1$.

Lastly, the third set of simulations takes a look at the role of the intercept of the conditional shock distribution, $\mu_{\omega^{*}}$, in particular, its interaction with $\sigma_{u}$.  We can think of $\mu_{\omega^{*}}$ as being the sole component of $\x$ that has no variance, i.e. because it's corresponding entry in the vector $\x$ is the scalar 1.  For larger values of $\mu_{\omega^{*}}$, the outperformance of the adjusted forecast declines with increases in $\sigma_{u}$, but this effect is absent for smaller values of $\mu_{\omega^{*}}$ 

\clearpage 

\section{Real Data Example}\label{Real Data Example}

We show the applicability of our method using a real data example that sits at the crossroads of financial trading and electoral politics.  In the spring of 2016 in the United States, the Republican Party's primary election process narrowed down candidates until Donald J. Trump cleared the threshold of votes to win the nomination formally at the party's convention that summer.  He would go on to face the Democratic Party's nominee, Hillary Rodham Clinton.   

From an ex-ante perspective, several qualities of the 2016 US election cycle as well as the candidates themselves made the election difficult to prognosticate.  The Electoral College permits victory without a majority or even plurality of the popular vote, which can render presidential races more competitive than a raw vote total would, elevating the uncertainty surrounding the country's future leadership.  The election featured no incumbent, ruling out any incumbent-advantage of the empirical, ``statistical" kind distinguished by \citet{mayhew2008incumbency}.  The Republican Party candidate espoused unorthodox, populist positions on matters such as healthcare, trade, and foreign policy, some of which could be considered rare in either of the major two parties.  Additionally, Donald J. Trump, lacking any experience in government --- either electoral or appointed service --- possessed neither a voting record nor any on-the-job performance for voters to judge or his opponents to attack. As one financial industry professional commented, comparing the 2016 election to the upcoming 2024 election, ``this time the markets will be aware of both possibilities and price them to some extent — we wouldn’t expect the same volatility as we saw in 2016 after the election" \citep{News_2024}. Gleaning signals from financial options markets and betting markets, \citet{wolfers2016financial} predicted that markets would decline prodigiously following a Trump victory in November 2016.  Finally, the election outcome delivered significant ``news", in the econometric sense of the word, in the simple sense that it was not predicted.  \citet{goodell2013us} found support for the theory that the polling-implied probabilities of election outcomes encode information about future macroeconomic conditions, which is itself reflected in market volatility.  In its final post before the election result, acclaimed forecasting outfit 538, headed by economist Nate Silver, predicted a Clinton victory with a probability of .714, more than 2-to-1 odds \citep{Silver_2016}, suggesting that Trump's victory was at least somewhat surprising.  The lack of uncertainty exerts downward pressure on volatility \citep{li2006presidential,bowes2018stock}, setting the table for a surprise.

For all of these reasons and more, the aftermath of the 2016 presidential election meets the standard of an interesting and notable event for which a quantitative researcher might seek a volatility point prediction.  On a more technical level, the election outcome was not known until the evening of election day, well after the closing of financial markets at 4pm Eastern Time.  This satisfies the condition that the shock be not yet digested by liquid markets.  We therefore proceed to make the following technical specifications in order to predict the volatility of financial services ETF IYG\footnote{It has been noted that GARCH effects are more attenuated in aggregated returns \citep{zivot2009practical}, which suggests against using the S\&P 500 or similar indices as an example.} (an ETF composed of American financial majors JPMorgan, Bank of American, etcetera) on Wednesday November 9th, 2016.

\begin{enumerate}
    \item \textbf{Model choice} We assume a GARCH(1,1) for the daily log return series of IYG in each donor.  As argued in \citet{hansen2005forecast}, a GARCH(1,1) is rarely dominated by more heavily-parameterized GARCH specifications.  It thus provides a defensible choice when motivation or time for choosing another model is lacking.  For the time series under study and the donor series alike, we fit a GARCH(1,1) on almost four years of market data prior to the shock.

    \item \textbf{Covariate Choice} We choose covariates that could plausibly satisfy the model assumptions spelled out earlier, that is, risk-related and macroeconomic covariates that could plausibly be weighted and summed in a shock distribution.  We thus choose the log return Crude Oil (CL.F), the VIX (VIX) and the log return of the VIX, the log returns of the 3-month, 5-year, 10-year, and 30-year US Treasuries, as well as the log return of the most recently available monthly spread between AAA and BAA corporate debt, widely considered a proxy for lending risk \citep{goodell2013us, kane1996p}.  We also include the log return in the trading volume of the ETF IYG itself, which serves as a proxy for panic.  Finally, we include the squares of the demeaned log return of IYG for the 30 trading days preceding the shocks.  For each variable in the volatility profile, we compute the sample mean and sample standard deviation across the $n+1$ events, allowing us to scale the variables to have zero mean and unit variance.  Hence, no single variable can dominate the distance-based weighting procedure.

    \item \textbf{Donor pool construction} We choose the three most recent US presidential elections prior to the 2016 election.  The three US presidential elections are the only presidential elections since the advent of the ETF IYG.  We exclude the midterm congressional elections in the US (i.e. those held in even years not divisible by four), which generate far lower voter turnout and feature no national races.

    \item \textbf{Choice of estimator for volatility} We use the sum of squared 5-minute log returns of IYG on November 9th, 2016, otherwise known as the Realized Volatility estimator of volatility \citep{andersen2008realized}, as our proxy.  We exclude the first five minutes of the trading day, resulting in a sum of 77 squared five-minute returns generated between 9:35am and 4pm.
    \item \textbf{Data Sources} All daily market data is provided via the YahooFinance API available in the quantmod package in R \citep{ryan2015package}.  In order to calculate log equity returns, we use Adjusted Close of IYG.  The realized volatility is computed using high-frequency quote data available from Wharton Research Data Services (WRDS) \citep{wachowicz2020wharton}.  The spread between is AAA and BAA yields is provided by Federal Reserve Economic Data (FRED) and accessed via the quantmod package.
\end{enumerate} 

\begin{figure}[H]
\begin{center}
  \textbf{Real Data Example: 2016 Election}\par\medskip
  \includegraphics[scale=.5]{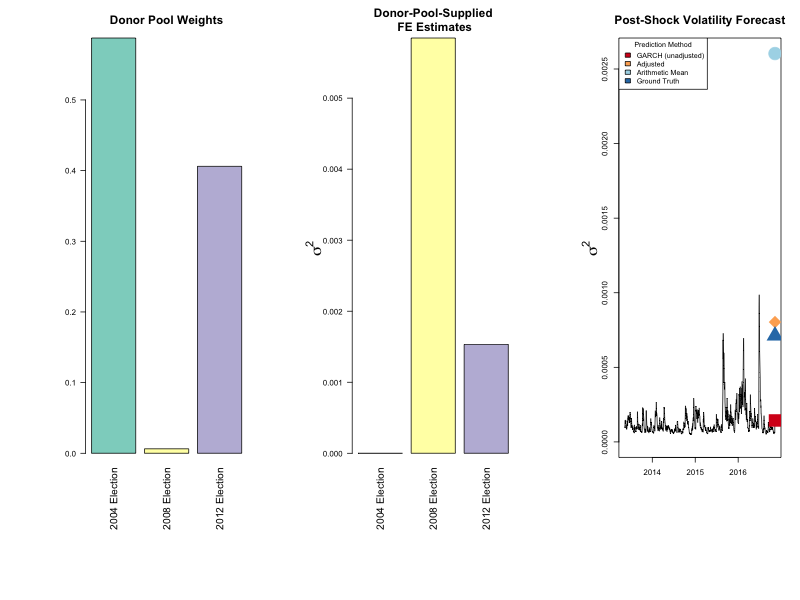}
  \caption{The volatility induced by the 2016 US election}
  \label{fig:SVF_2016}
  \end{center}
\end{figure}

We now discuss the three subplots in Figure \ref{fig:SVF_2016} in order from left to right.  On the left, we see that distanced-based weighting places nearly equal weight on the 2004 and 2012 elections, with only neglible weight on the 2008 election, when financial market conditions were extreme across nearly every dimension.  Assuming an approximately correct specification of the covariates, this is interpreted to mean that even of the 2016 US election had a general climate of risk and tension less extreme than 2008 and more similar to the 2004 and 2012 elections.  In the middle plot, we notice that the fixed effect estimates for 2008 and 2012 are considerable, with only 2004 registering a near-zero fixed effect estimate.  The fixed effect estimates quantify the amount of surprise the US election results delivered (strictly speaking, not only the presidential race but all November elections in the US with the ability to influence financial markets) under the assumption of a GARCH(1,1).  As estimates gleaned from only one data point per time series, they are theoretically high in variance.  On the right, we observe in black the  $\hat\sigma^{2}$ yielded by the GARCH(1,1) for the time series under study.  We also observe four colored points, each listen in the legend: three predictions and the ground truth.  We include the prediction derived by adjusting the GARCH(1,1) prediction by the arithmetic mean of the fixed effect estimates.  As is evident, our method comes reasonably close the ground truth.  The prediction is not only directionally correct, i.e. we predict a volatility spike where there is one; the prediction far outperforms the unadjusted prediction.  Remarkably, the arithmetic-mean based prediction here demonstrates the inherent risk in failing to weight each donor appropriately.  The 2008 election receives far more weight than is called for, as simple averaging ignores the radically different conditons on the evening of those two events.  

Naturally, one might ask how sensitive this prediction is to at least two kinds of model specification: donor pool specification and covariate specification.  There are two responses to these concerns.  First, although the practitioner lacks a priori knowledge of the adequacy of the donors with respect to the time series under study, it is possible to gauge the diversity of the donor information by examining the singular values of the volatility profile.  In the prediction presented here, the singular values descend from 62$\%$ to 23$\%$ to 15$\%$ of the the cumulative variation, indicating a moderate concentration or redundancy of information in the three donors.  Second, we follow \citet{steegen2016increasing} in a executing a multiverse analysis.  In particular, in the supplement, we carry out leave-one-out analyses on both the donor set and the covariate set.  Our multiverse analysis concludes that our adjustment approach beats the unadjusted forecast in every leave-one-out configuration. These results include an added donor representing Brexit.

\section{Discussion}

This present work applies and innovates techniques in similarity-based forecasting and information aggregation in order to provide better GARCH forecasts under news shocks.  It extends \citet{lin2021minimizing} principally by substituting the GARCH model for an AR(1), i.e. by modeling both the mean and volatility of a univariate time series.  An interesting connection insight related to \citet{lin2021minimizing} is that the GARCH model, under weak assumptions, can be represented as an ARMA on the squared residuals $a^{2}_{t}$.  Hence, a shock $\omega^{*}_{i,t}$ to the volatility at time $t$ is an identical shock to $a^{2}_{t}$. However, this use-case becomes less attractive in situations where the sign of the time series under study following the shock is uncertain.  Insofar as this work has made advances that accommodate heteroskedastic time series, the benefits may redound most amply to applications like Value-At-Risk (VaR) and Expected Shortfall, where $\sigma^{2}_{t}$ is an input.

This work as well as \citet{lin2021minimizing} itself can be viewed as a generalization of \cite{phillips1996forecasting} where the rare event probability $\lambda$ is assumed to be 1 and the error associated with the rare event is estimated using external series.  In our setting, if the shock to the time series under study were known and yet so underdescribed, i.e. so lacking in qualitative context, then \cite{phillips1996forecasting} would be recommended.  However, in our setting the rare event is not only rare but also contains specific qualitative features that are best accounted for via data aggregation.

The method under development does not strictly require knowledge of the length of the shocks in the donor pool, but correctly sizing up those shock lengths is helpful to proper estimation of the shocks in the donor pool.  An important question remains: even if the donor pool shock lengths are assumed to be known, how do we advise the operator to forecast the time series under study?  In other words, for how long is the adjustment estimator $\omega^{*}$ valid, applicable, and reliable?  One idea suggested by the paper is obvious: why not aggregate the shock lengths from the donors as well and round that quantity or take the floor or ceiling of any non-integer value?  This is worth pursuing.  However, it may be that estimating the persistence of a volatility shock induced by news is an endeavor deserving of its own study, where aggregation methods might naturally arise as helpful tools.

There is also a broader discusion to be had regarding the degree of model heterogeneity permitted in fitting the donors' series.  

\subsection{Comparison with KNN}

\cite{clements1996intercept} (cited in \cite{guerron2017macroeconomic})  note intercept corrections find their theoretical basis in the occurrence of structural breaks, whereas Nearest-Neighbor methods, being nonparametric, are theoretically more adept at accounting for nonlinearity.  The present work examines news shocks, which are more closely related to structural breaks.  Hence, neither nearest neighbor methods nor nonlinearity figure heavily in our work.  However, there are deeper observations to be made about KNN as it relates to our method.

The method presented here is unlike traditional KNN in that we are not trying to learn a function, first and foremost.  We are trying to estimate a parameter.  KNN runs into the problem: the curse of dimensionality.  In contrast, large $p$ is not a problem in synthetic methods, because the thing estimated is the vector $w$ with $n-1$ degrees of freedom.  For KNN, a high-dimensional space, i.e. large $p$, corresponding to many covariates, is a difficult space in which to work with distances \citep{hastie2009elements}.  In contrast, large $p$ is not a problem in and of itself for synthetic control --- in fact, asymptotic results exist for $p$ \citep{abadie2010synthetic}.  

As is pointed out in \citet{hastie2009elements}, KNN regression performs well when $K$ is chosen small enough that one can simply average the points $\{y_{i}\}_{i=1}^{N_{train}}$in the neighborhood around each element in $\{y_{i}\}_{i=1}^{N_{test}}$ to get good predictions.  As we have noted above, the arithmetic mean-based estimator of $\omega^{*}$, denoted $\overline{\omega^{*}}$, corresponds to KNN when $K = n$, the number of donors.  Fundamentally, the idea that $n$ is small enough and the donors are homogeneous enough that one could simply average the $\hat\omega_{i}$ is at odds with the assumed variation in the shock effects.

In KNN regression, the hyperparameter K must be learned.  In similarity-based parameter correction, the number of donors is not learned.  A donor pool is curated, and then careful rules of thumb can be applied to determine whether a given donor should be included or excluded.  While it would not necessarily hurt to `learn' the appropriate number of donors to use, this information would probably not be as useful as knowing which donors and covariates provide the basis for the best forecasts.  This brings us to a deeper point about the distinction between similarity-based methods in the style of \citet{lin2021minimizing} and KNN.  In KNN, the exogenous variables are taken as a given and nearness to the object to be predicted depends on the distance function chosen.  In contrast, in \citet{lin2021minimizing}, the determination of nearness begins with a qualitative step, i.e. curating the units between which we will calculate distances and from we will ultimately derive weights.

\subsection{Donor Pool Construction}

Should we gather as many donors as possible and pick them quantitatively?  It would be counter to the method proposed to assemble a vast number of donors, lacking careful scrutiny of the qualitative fit, and let the optimization simply pick the donors, via weighting.  What makes a donor good is not merely its quantitative fit but its qualitative fit as well.  \citet{abadie2022synthetic} make a similar point about large donor pools.  What matters is that the donors chosen are properly situated in the $p$-dimensional predictor space, so as to allow proper estimation of the weights.  For more on this question, see the Supplement, where the donors and the volatility profile are treated with a leave-one-out analysis.

\subsection{The Nature and Estimation of Volatility Shocks}

Not all of the volatility of an asset return may be related to news \citep{boudoukh2019information}.  This explains our inclusion of an idiosyncratic noise term in the shock specification.  However, this point also gestures in the direction of possible unexplained variation in the shocks.  \citet{chinco2019sparse} find that for predicting 1-minute returns, highly transitory firm-specific news is useful.  The authors conclude that news about fundamentals is predictive.

It would be a pyrrhic victory for our method if the volatility profile indeed underlies real-world shocks but the volatility profile is radically high-dimensional or the signal in the shocks is overwhelmed by the noise term.  Even unbiased predictions can be unhelpful if they are high in variability, and in Section \ref{forecastcomb}, we show how the benefits of forecast combination.  Another possibility is high-frequency data and the use of linear models like HAR \citep{corsi2012har}.  HAR would not only increase the sample size available for discovering the autoregressive structure of a series' realized volatility.  It would also open the door to high-dimensional regression methods, shrinkage estimators, and more.

\subsection{Parameter Stability}

There is an important question about the stability of the GARCH parameters under the presence of a shock, and on parameter instability as it pertains to forecasting, we refer readers to \cite{rossi2013advances, rossi2021forecasting}.  There are at least two reasons that we do not explore parameter instability or methods to adjust for it.  First, the marginal effect of coefficient changes at the shock time would, under the assumptions in this work, be swamped by the random effect.  Second, the estimation of post-shock parameter values would require at least several --- better yet, dozens --- of post-shock data points, whereas this work assumes access to zero post-shock data points.  However, it is possible that similarity-based estimators for the GARCH coefficients could be produced, for example, by adapting the methods of \citet{dendramis2020similarity}.

\section{Supplement}

\subsection{Leave-one-out: How we analyzed a multiverse of 50 predictions}

Given the option of redoing our analysis, with nine covariates and four donors, a natural question from a place of skepticism is, how stable are the results --- how contingent are they on a particular specification?  To answer these questions, in Table \ref{tab:prediction_table_with Brexit as June 22nd, 2016} we generate all fifty predictions yielded the decision of leave out any one of the donors (or none) and any one of the covariates (or none).  This analysis includes the prediction presented above, which can be viewed as the null model, at least in the sense that it provides and baseline for comparison.

\begin{table}[ht]
  \centering
  \captionof{table}{Adjusted Forecasts Ranked by QL Loss with Brexit as June 22nd, 2016} \label{tab:prediction_table_with Brexit as June 22nd, 2016} 
  \begingroup\fontsize{7pt}{8pt}\selectfont
  \begin{tabular}{ccc}
    
    \hline
 QL Loss of the Adjusted Forecast& Omitted Covariate & Omitted Donor \\ 
    \hline
    0.0003 & Log Return of VIX & 2016-06-22 \\ 
    0.0013 & Log Return of IRX & 2008-11-04 \\ 
    0.0013 & Log Return of TYX & None \\ 
    0.0013 & Log Return of TYX & 2016-06-22 \\ 
    0.0017 & Log Return of TYX & 2008-11-04 \\ 
    0.0019 & Log Return of TNX & 2016-06-22 \\ 
    0.0020 & Log Return of TNX & 2008-11-04 \\ 
    0.0027 & Log Return of FVX & 2008-11-04 \\ 
    0.0029 & VIX & 2008-11-04 \\ 
    0.0034 & Log Return of TNX & None \\ 
    0.0036 & Log Return of CL.F & 2016-06-22 \\ 
    0.0037 & Log Return of VIX & 2008-11-04 \\ 
    0.0041 & Debt Risk Spread & 2016-06-22 \\ 
    0.0043 & None & 2008-11-04 \\ 
    0.0047 & Log Return of FVX & 2016-06-22 \\ 
    \rowcolor{gray} 0.005 & NA (Median Forecast) & NA (Median Forecast)\\  
    0.0053 & VIX & 2016-06-22 \\ 
    0.0062 & Log Return of CL.F & 2008-11-04 \\
    \rowcolor{gray} 0.0068 & NA (Average Forecast) & NA (Average Forecast)\\ 
    \rowcolor{yellow} 0.0070 & None & 2016-06-22 \\ 
    0.0075 & Log Return of FVX & None \\ 
    0.0077 & Log Return of IRX & 2016-06-22 \\ 
    0.0102 & Log Return of IRX & None \\ 
    0.0103 & Debt Risk Spread & 2008-11-04 \\ 
    0.0113 & Log Return of VIX & None \\ 
    0.0135 & VIX & None \\ 
    0.0135 & IYG & 2008-11-04 \\ 
    \rowcolor{yellow} 0.0136 & None & None \\ 
    0.0137 & Log Return of CL.F & None \\ 
    0.0154 & Debt Risk Spread & None \\ 
    0.0268 & IYG & 2016-06-22 \\
     
    0.0568 & Log Return of TNX & 2012-11-06 \\ 
    0.0670 & IYG & None \\ 
    0.0748 & Log Return of VIX & 2012-11-06 \\ 
    0.0805 & Log Return of TNX & 2004-11-02 \\ 
    0.0808 & Log Return of TYX & 2004-11-02 \\ 
    0.0816 & Log Return of VIX & 2004-11-02 \\ 
    0.0851 & Debt Risk Spread & 2012-11-06 \\ 
    0.0864 & None & 2012-11-06 \\ 
    0.0872 & VIX & 2012-11-06 \\ 
    0.0899 & Log Return of IRX & 2004-11-02 \\ 
    0.0900 & Log Return of CL.F & 2012-11-06 \\ 
    0.0907 & Log Return of IRX & 2012-11-06 \\ 
    0.0913 & VIX & 2004-11-02 \\ 
    0.0916 & Log Return of FVX & 2004-11-02 \\ 
    0.0917 & None & 2004-11-02 \\ 
    0.0925 & Debt Risk Spread & 2004-11-02 \\ 
    0.0935 & Log Return of CL.F & 2004-11-02 \\ 
    0.0993 & Log Return of TYX & 2012-11-06 \\ 
    0.1142 & Log Return of FVX & 2012-11-06 \\ 
    0.2143 & IYG & 2004-11-02 \\ 
    2.3529 & IYG & 2012-11-06 \\ 
    \rowcolor{red} 2.3529 & All & All \\
     \hline
  \end{tabular}
  
  \endgroup
  \end{table}

  \begin{table}[ht]
    \centering
    \captionof{table}{Adjusted Forecasts Ranked by QL Loss with Brexit as June 23rd, 2016} \label{tab:prediction_table_with Brexit as June 23rd, 2016} 
    \begingroup\fontsize{7pt}{8pt}\selectfont
    \begin{tabular}{ccc}
      
      \hline
   QL Loss of the Adjusted Forecast& Omitted Covariate & Omitted Donor \\ 
      \hline
      0.0003 & Log Return of VIX & 2016-06-23 \\ 
  0.0013 & Log Return of TYX & 2016-06-23 \\ 
  0.0019 & Log Return of TNX & 2016-06-23 \\ 
  0.0036 & Log Return of CL.F & 2016-06-23 \\ 
  0.0041 & Debt Risk Spread & 2016-06-23 \\ 
  0.0047 & Log Return of FVX & 2016-06-23 \\ 
  0.0054 & VIX & 2016-06-23 \\ 
  \rowcolor{yellow}0.0070 & None & 2016-06-23 \\ 
  0.0079 & Log Return of IRX & 2016-06-23 \\ 
  0.0268 & IYG & 2016-06-23 \\ 
  0.0362 & IYG & None \\ 
  0.1113 & IYG & 2008-11-04 \\ 
  0.2401 & Log Return of TNX & 2012-11-06 \\ 
  0.2420 & Log Return of FVX & 2012-11-06 \\ 
  0.2553 & Log Return of TNX & None \\ 
  0.2678 & IYG & 2012-11-06 \\ 
  0.2717 & Log Return of FVX & None \\ 
  0.2725 & Log Return of TYX & None \\ 
  0.2766 & Log Return of CL.F & None \\ 
  0.2881 & VIX & None \\ 
  0.2884 & Debt Risk Spread & None \\ 
  \rowcolor{yellow} 0.2905 & None & None \\ 
  0.2954 & Log Return of IRX & None \\ 
  0.2964 & Log Return of TYX & 2012-11-06 \\ 
  0.3086 & Log Return of IRX & 2012-11-06 \\ 
  \rowcolor{gray}0.3125 & NA (Median Forecast) & NA (Median Forecast)\\
  0.3164 & VIX & 2012-11-06 \\ 
  0.3186 & Log Return of CL.F & 2012-11-06 \\ 
  0.3219 & None & 2012-11-06 \\ 
  0.3257 & Debt Risk Spread & 2012-11-06 \\ 
  0.3289 & VIX & 2008-11-04 \\ 
  0.3344 & Log Return of CL.F & 2008-11-04 \\ 
  \rowcolor{gray}0.3125 & NA (Average Forecast) & NA (Average Forecast)\\
  0.3475 & Log Return of VIX & 2012-11-06 \\ 
  0.3581 & Log Return of IRX & 2008-11-04 \\ 
  0.3586 & None & 2008-11-04 \\ 
  0.3617 & Log Return of FVX & 2008-11-04 \\ 
  0.3619 & Debt Risk Spread & 2008-11-04 \\ 
  0.3642 & Log Return of TNX & 2008-11-04 \\ 
  0.3695 & Log Return of TYX & 2008-11-04 \\ 
  0.3801 & Log Return of VIX & None \\ 
  0.3902 & Log Return of VIX & 2008-11-04 \\ 
  0.5510 & IYG & 2004-11-02 \\ 
  0.7452 & Log Return of FVX & 2004-11-02 \\ 
  0.7532 & Log Return of TNX & 2004-11-02 \\ 
  0.7591 & Log Return of CL.F & 2004-11-02 \\ 
  0.7616 & Debt Risk Spread & 2004-11-02 \\ 
  0.7681 & None & 2004-11-02 \\ 
  0.7683 & VIX & 2004-11-02 \\ 
  0.7753 & Log Return of IRX & 2004-11-02 \\ 
  0.8034 & Log Return of VIX & 2004-11-02 \\ 
  0.8618 & Log Return of TYX & 2004-11-02 \\ 
      \rowcolor{red} 2.3529 & All & All \\
       \hline
    \end{tabular}
    
    \endgroup
    \end{table}

\subsection{Sensitivity to Covariates Chosen}
Any covariate that appears more often closer to the bottom of Table \ref{tab:prediction_table_with Brexit as June 22nd, 2016} is, $\textit{ceteris paribus}$, a more important covariate for this prediction task, since by dropping it, a larger loss results.  The covariates that thus stand out are the demeaned log return of IYG, the VIX, the Debt Risk Spread (spread between between AAA and BAA corporate debt), and last but not least, the choice of dropping none of the covariates.  The demeaned log return of IYG, of which we use the 30 days preceeding the $T_{i}^{*}$, suggests an unusually strong relevance in matching donors to the time series under study.  The Debt Risk Spread is also deserving of additional comment and attention, perhaps alongside the poor performance of the log return of the VIX.  The Debt Risk Spread is available at a monthly frequency via FRED (add citation), whereas the VIX is available daily via Yahoo Finance.  It may be the case that the true conditional shock distribution includes low-frequency data as well as data in levels (like the VIX), while the daily changes in the VIX are absent.

\subsection{Sensitivity to Donors Chosen}

The most glaring result visible in Table \ref{tab:prediction_table_with Brexit as June 22nd, 2016} regarding donor selection is the poor performance of the 2008 US Election.  In hindsight, this is unsurprising, given the large estimated fixed effect for November 5th, 2008 as well as the swirl of complex events occurring in financial markets around that time.  It is possible that the GARCH(1,1) used to fit the run-up to that election is underparameterized.  The weight given to the 2008 US election is near-zero, but in our method, unsuitable donors can influence the distance-based weighting by affecting that dispersion of one or more of the $p$ covariates present in the volatility profile.  Recall that for each of the $p$ covariates, we transform the $n+1$-vector of variables to Z-scores.  The lesson here is that more careful pre-quantitative inspection of the donors may be warranted as well as ordinary outlier analysis for the volatility profile.  As an additional note, dropping the 2008 Election vastly improves the performance of the arithmetic mean forecast.  This provides a cautionary tale against using the arithmetic mean estimator for the conditional shock effect.  

The donors most conducive to good predictions in this task are the 2004 and 2008 US elections.  The estimated fixed effects tell a story of only modest surprise due to these election results: the estimate is nearly zero for 2004, while 2012 is nonzero but small.  Given that the presidential election results of these two did not deliver much surprise, it is possible that the surprise could be due more so to non-presidential races.

We discuss one last remarkable phenomenon.  The prediction that dropped the preceding 30 days of demeaned log returns of IYG as well as the 2012 Election yielded the same QL Loss as the unadjusted forecast.  By inspecting the donor weight and donor fixed effect estimates, we can see that nearly zero adjustment was made because the 2004 Election is given nearly all the weight.

\subsubsection{Brexit}
In the pre-quantitative analysis of donors, there was only one competitor to the setup that was ultimately selected and presented in Section \ref{Real Data Example}, and that competitor donor pool used Brexit.  The inclusion of Brexit is based on the fundamental belief that the conditional shock distribution governing IYG's volatility shocks may be shared among US elections and some political events elsewhere, like the Brexit referendum of July 23rd, 2016.  That referendum shared some important qualities with the 2016 US elections, which we have discussed above.

We now turn, however, to why the inclusion of Brexit failed to perform nearly as well our primary specification.  First, note time zone differences between the UK and the US make difficult the analysis of after-hour shocks on US markets.  Additionally, like many election results, that the ``Leave'' side of the referendum would prevail was not revealed at any discrete time, of course, but became more certain as observation of voting turnout and informal vote tallies across the UK progressed \citep{BBC_News_2016}.  This is in spite of UK news outlets adhering to various rules and guidelines regarding reportage of polling results \citep{Bailey_2024}.  These facts pose a challenge for any method that uses daily data.  Of prime concern is whether we should consider $T^{*}$ for Brexit to have occurred after market hours on June 23rd, 2016, Eastern Standard Time, which ignores the steady percolation of information and uncertainty that attend election days, or alternatively, consider the shock to have occurred after hours on June 22nd, 2016, which implies that $T^{*}+1$ is June 23rd, 2016, and hence the reaction of IYG to Brexit can be estimated and extracted from late-day trading on that date.  Given this dilemma, we opt for dropping Brexit from our predictive model completely.

\subsection{Can we combine forecasts to outperform the individual forecasts?}\label{forecastcomb}

Forecast combination is technique with a vast, sprawling literature that we have referred to above.  While forecast combination can be justified in any number of ways, here we invoke forecast combination as a way to robustify forecasts against misspecification.  We combine in two simply ways, using the mean of all 50 forecasts and the median of all 50 forecasts.  The results are presented in Tables \ref{tab:prediction_table_with Brexit as June 22nd, 2016} and \ref{tab:prediction_table_with Brexit as June 23rd, 2016}.

\begin{figure}[H]
  \begin{center}
    \textbf{Real Data Example: 2016 Election with Brexit as June 22nd, 2016}\par\medskip
    \includegraphics[scale=.6]{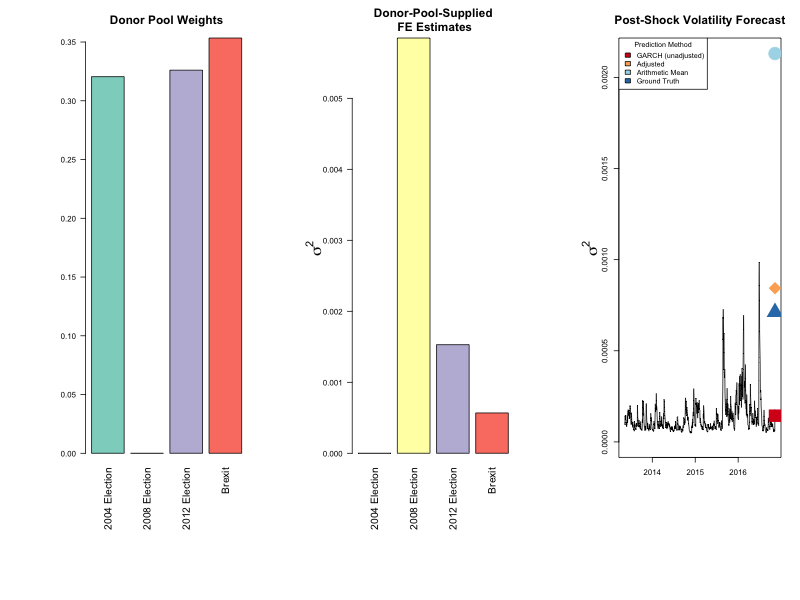}
    \caption{When Brexit is taken to be June 22nd, 2016, it is shares equal weight with the 2004 and 2008 elections, while its modest fixed effect estimate is the smallest of the three nonzero estimates.}
    \label{fig:SVF_2016_with_Brexit}
    \end{center}
  \end{figure}

  \begin{figure}[H]
    \begin{center}
      \textbf{Real Data Example: 2016 Election with an Brexit as June 23rd, 2016}\par\medskip
      \includegraphics[scale=.6]{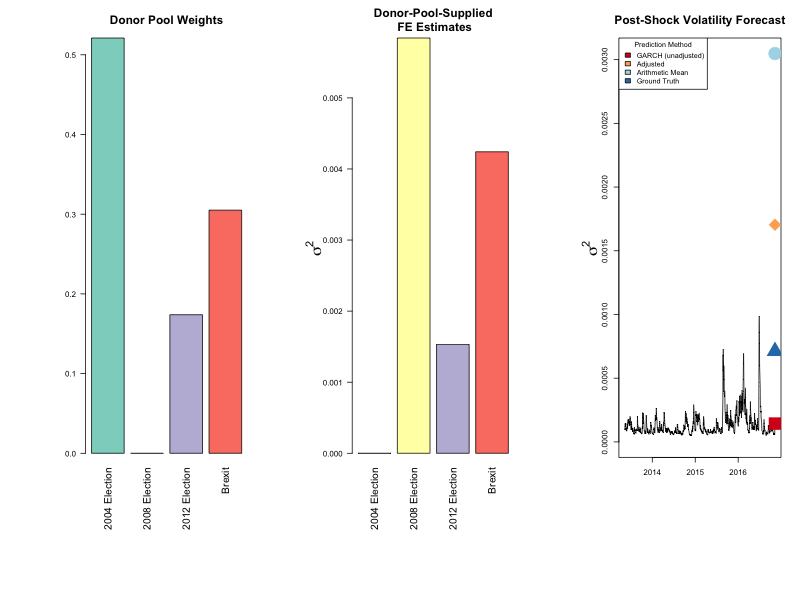}
      \caption{When Brexit is taken to be June 23rd, 2016, its own weight is unchanged but others' weights are.  The large fixed effect estimate of Brexit contributes to a substantial overprediction.}
      \label{fig:SVF_2016_with_Brexit}
      \end{center}
    \end{figure}
  
\clearpage

\bibliography{synthVolForecast}

\begin{thebibliography}{58}
\providecommand{\natexlab}[1]{#1}
\providecommand{\url}[1]{\texttt{#1}}
\expandafter\ifx\csname urlstyle\endcsname\relax
  \providecommand{\doi}[1]{doi: #1}\else
  \providecommand{\doi}{doi: \begingroup \urlstyle{rm}\Url}\fi

\bibitem[Abadie and Gardeazabal(2003)]{abadie2003economic}
Alberto Abadie and Javier Gardeazabal.
\newblock The economic costs of conflict: A case study of the basque country.
\newblock \emph{American Economic Review}, 93\penalty0 (1):\penalty0 113--132, 2003.

\bibitem[Abadie and Vives-i Bastida(2022)]{abadie2022synthetic}
Alberto Abadie and Jaume Vives-i Bastida.
\newblock Synthetic controls in action.
\newblock \emph{arXiv preprint arXiv:2203.06279}, 2022.

\bibitem[Abadie et~al.(2010)Abadie, Diamond, and Hainmueller]{abadie2010synthetic}
Alberto Abadie, Alexis Diamond, and Jens Hainmueller.
\newblock Synthetic control methods for comparative case studies: Estimating the effect of california’s tobacco control program.
\newblock \emph{Journal of the American Statistical Association}, 105\penalty0 (490):\penalty0 493--505, 2010.

\bibitem[Andersen and Benzoni(2010)]{andersen2010stochastic}
Torben~G Andersen and Luca Benzoni.
\newblock Stochastic volatility.
\newblock \emph{CREATES Research Paper}, \penalty0 (2010-10), 2010.

\bibitem[Andersen et~al.(2003)Andersen, Bollerslev, Diebold, and Labys]{andersen2003modeling}
Torben~G Andersen, Tim Bollerslev, Francis~X Diebold, and Paul Labys.
\newblock Modeling and forecasting realized volatility.
\newblock \emph{Econometrica}, 71\penalty0 (2):\penalty0 579--625, 2003.

\bibitem[Andersen et~al.(2007)Andersen, Bollerslev, and Diebold]{andersen2007roughing}
Torben~G Andersen, Tim Bollerslev, and Francis~X Diebold.
\newblock Roughing it up: Including jump components in the measurement, modeling, and forecasting of return volatility.
\newblock \emph{The review of economics and statistics}, 89\penalty0 (4):\penalty0 701--720, 2007.

\bibitem[Andersen and Benzoni(2008)]{andersen2008realized}
Torben~Gustav Andersen and Luca Benzoni.
\newblock Realized volatility, working paper 2008-14.
\newblock 2008.

\bibitem[Bailey(2024)]{Bailey_2024}
Ric Bailey.
\newblock How the bbc reports polling day, May 2024.
\newblock URL \url{https://www.bbc.com/news/uk-politics-48124106}.

\bibitem[Bauwens et~al.(2006)Bauwens, Preminger, and Rombouts]{bauwens2006regime}
Luc Bauwens, Arie Preminger, and Jeroen~VK Rombouts.
\newblock Regime switching garch models.
\newblock 2006.

\bibitem[Bollerslev(1986)]{bollerslev1986generalized}
Tim Bollerslev.
\newblock Generalized autoregressive conditional heteroskedasticity.
\newblock \emph{Journal of econometrics}, 31\penalty0 (3):\penalty0 307--327, 1986.

\bibitem[Boudoukh et~al.(2019)Boudoukh, Feldman, Kogan, and Richardson]{boudoukh2019information}
Jacob Boudoukh, Ronen Feldman, Shimon Kogan, and Matthew Richardson.
\newblock Information, trading, and volatility: Evidence from firm-specific news.
\newblock \emph{The Review of Financial Studies}, 32\penalty0 (3):\penalty0 992--1033, 2019.

\bibitem[Bowes(2018)]{bowes2018stock}
David~R Bowes.
\newblock Stock market volatility and presidential election uncertainty: Evidence from political futures markets.
\newblock \emph{Journal of Applied Business Research}, 34\penalty0 (1), 2018.

\bibitem[Box(2013)]{box2013box}
George Box.
\newblock Box and jenkins: time series analysis, forecasting and control.
\newblock In \emph{A Very British Affair: Six Britons and the Development of Time Series Analysis During the 20th Century}, pages 161--215. Springer, 2013.

\bibitem[Brownlees et~al.(2011)Brownlees, Engle, and Kelly]{brownlees2011practical}
Christian~T Brownlees, Robert~F Engle, and Bryan~T Kelly.
\newblock A practical guide to volatility forecasting through calm and storm.
\newblock \emph{Available at SSRN 1502915}, 2011.

\bibitem[Chinco et~al.(2019)Chinco, Clark-Joseph, and Ye]{chinco2019sparse}
Alex Chinco, Adam~D Clark-Joseph, and Mao Ye.
\newblock Sparse signals in the cross-section of returns.
\newblock \emph{The Journal of Finance}, 74\penalty0 (1):\penalty0 449--492, 2019.

\bibitem[Christensen and Prabhala(1998)]{christensen1998relation}
Bent~J Christensen and Nagpurnanand~R Prabhala.
\newblock The relation between implied and realized volatility.
\newblock \emph{Journal of Financial Economics}, 50\penalty0 (2):\penalty0 125--150, 1998.

\bibitem[Clements and Hendry(1998)]{clements1998forecasting}
Michael Clements and David~F Hendry.
\newblock \emph{Forecasting economic time series}.
\newblock Cambridge University Press, 1998.

\bibitem[Clements and Hendry(1996)]{clements1996intercept}
Michael~P Clements and David~F Hendry.
\newblock Intercept corrections and structural change.
\newblock \emph{Journal of Applied Econometrics}, 11\penalty0 (5):\penalty0 475--494, 1996.

\bibitem[Corsi et~al.(2012)Corsi, Audrino, and Ren{\'o}]{corsi2012har}
Fulvio Corsi, Francesco Audrino, and Roberto Ren{\'o}.
\newblock Har modeling for realized volatility forecasting.
\newblock 2012.

\bibitem[De~Luca et~al.(2006)]{de2006forecasting}
Giovanni De~Luca et~al.
\newblock Forecasting volatility using high-frequency data.
\newblock \emph{Statistica Applicata}, 18, 2006.

\bibitem[Dendramis et~al.(2020)Dendramis, Kapetanios, and Marcellino]{dendramis2020similarity}
Yiannis Dendramis, George Kapetanios, and Massimiliano Marcellino.
\newblock A similarity-based approach for macroeconomic forecasting.
\newblock \emph{Journal of the Royal Statistical Society Series A: Statistics in Society}, 183\penalty0 (3):\penalty0 801--827, 2020.

\bibitem[Dominguez and Panthaki(2006)]{dominguez2006defines}
Kathryn~ME Dominguez and Freyan Panthaki.
\newblock What defines ‘news’ in foreign exchange markets?
\newblock \emph{Journal of International Money and Finance}, 25\penalty0 (1):\penalty0 168--198, 2006.

\bibitem[Engle(1982)]{engle1982autoregressive}
Robert~F Engle.
\newblock Autoregressive conditional heteroscedasticity with estimates of the variance of united kingdom inflation.
\newblock \emph{Econometrica: Journal of the econometric society}, pages 987--1007, 1982.

\bibitem[Engle and Patton(2001)]{engle2001good}
Robert~F Engle and Andrew~J Patton.
\newblock What good is a volatility model?
\newblock \emph{Quantitative finance}, 1\penalty0 (2):\penalty0 237, 2001.

\bibitem[Foroni et~al.(2022)Foroni, Marcellino, and Stevanovic]{foroni2022forecasting}
Claudia Foroni, Massimiliano Marcellino, and Dalibor Stevanovic.
\newblock Forecasting the covid-19 recession and recovery: Lessons from the financial crisis.
\newblock \emph{International Journal of Forecasting}, 38\penalty0 (2):\penalty0 596--612, 2022.

\bibitem[Francq and Zakoian(2019)]{francq2019garch}
Christian Francq and Jean-Michel Zakoian.
\newblock \emph{GARCH models: structure, statistical inference and financial applications}.
\newblock John Wiley \& Sons, 2019.

\bibitem[Goodell and V{\"a}h{\"a}maa(2013)]{goodell2013us}
John~W Goodell and Sami V{\"a}h{\"a}maa.
\newblock Us presidential elections and implied volatility: The role of political uncertainty.
\newblock \emph{Journal of Banking \& Finance}, 37\penalty0 (3):\penalty0 1108--1117, 2013.

\bibitem[Guerr{\'o}n-Quintana and Zhong(2017)]{guerron2017macroeconomic}
Pablo Guerr{\'o}n-Quintana and Molin Zhong.
\newblock Macroeconomic forecasting in times of crises.
\newblock 2017.

\bibitem[Han and Kristensen(2014)]{han2014asymptotic}
Heejoon Han and Dennis Kristensen.
\newblock Asymptotic theory for the qmle in garch-x models with stationary and nonstationary covariates.
\newblock \emph{Journal of Business \& Economic Statistics}, 32\penalty0 (3):\penalty0 416--429, 2014.

\bibitem[Hansen and Lunde(2005)]{hansen2005forecast}
Peter~R Hansen and Asger Lunde.
\newblock A forecast comparison of volatility models: does anything beat a garch (1,1)?
\newblock \emph{Journal of applied econometrics}, 20\penalty0 (7):\penalty0 873--889, 2005.

\bibitem[Hansen et~al.(2012)Hansen, Huang, and Shek]{hansen2012realized}
Peter~Reinhard Hansen, Zhuo Huang, and Howard~Howan Shek.
\newblock Realized garch: a joint model for returns and realized measures of volatility.
\newblock \emph{Journal of Applied Econometrics}, 27\penalty0 (6):\penalty0 877--906, 2012.

\bibitem[Hastie et~al.(2009)Hastie, Tibshirani, and Friedman]{hastie2009elements}
Trevor Hastie, Robert Tibshirani, and Jerome Friedman.
\newblock \emph{The elements of statistical learning: data mining, inference, and prediction}.
\newblock Springer Science \& Business Media, 2009.

\bibitem[Hendry and Clements(1994)]{hendry1994theory}
D~Hendry and M~Clements.
\newblock On a theory of intercept corrections in macroeconometric forecasting.
\newblock 1994.

\bibitem[Kane et~al.(1996)Kane, Marcus, and Noh]{kane1996p}
Alex Kane, Alan~J Marcus, and Jaesun Noh.
\newblock The p/e multiple and market volatility.
\newblock \emph{Financial Analysts Journal}, 52\penalty0 (4):\penalty0 16--24, 1996.

\bibitem[Kenton(2022)]{Kenton}
Will Kenton.
\newblock Event-driven investing strategies and examples, 2022.
\newblock URL \url{https://www.investopedia.com/terms/e/eventdriven.asp}.

\bibitem[Kilian and L{\"u}tkepohl(2017)]{kilian2017structural}
Lutz Kilian and Helmut L{\"u}tkepohl.
\newblock \emph{Structural vector autoregressive analysis}.
\newblock Cambridge University Press, 2017.

\bibitem[Li and Born(2006)]{li2006presidential}
Jinliang Li and Jeffery~A Born.
\newblock Presidential election uncertainty and common stock returns in the united states.
\newblock \emph{Journal of Financial Research}, 29\penalty0 (4):\penalty0 609--622, 2006.

\bibitem[Lin and Eck(2021)]{lin2021minimizing}
Jilei Lin and Daniel~J Eck.
\newblock Minimizing post-shock forecasting error through aggregation of outside information.
\newblock \emph{International Journal of Forecasting}, 2021.

\bibitem[Mayhew(2008)]{mayhew2008incumbency}
David~R Mayhew.
\newblock Incumbency advantage in us presidential elections: The historical record.
\newblock \emph{Political Science Quarterly}, 123\penalty0 (2):\penalty0 201--228, 2008.

\bibitem[Mayhew(1995)]{mayhew1995implied}
Stewart Mayhew.
\newblock Implied volatility.
\newblock \emph{Financial Analysts Journal}, 51\penalty0 (4):\penalty0 8--20, 1995.

\bibitem[News(2024)]{News_2024}
Bloomberg News.
\newblock Election 2024: What will markets do with trump victory over biden?
\newblock \url{https://www.bloomberg.com/news/articles/2024-01-21/trump-s-2016-win-shook-markets-traders-won-t-get-fooled-again}, Jan 2024.

\bibitem[None(2016)]{BBC_News_2016}
None.
\newblock Leave campaign ahead in uk's eu referendum vote, Jun 2016.
\newblock URL \url{https://www.bbc.com/news/uk-politics-eu-referendum-36612368}.

\bibitem[Phillips(1996)]{phillips1996forecasting}
Robert~F Phillips.
\newblock Forecasting in the presence of large shocks.
\newblock \emph{Journal of Economic Dynamics and Control}, 20\penalty0 (9-10):\penalty0 1581--1608, 1996.

\bibitem[Romer and Romer(1989)]{romer1989does}
Christina~D Romer and David~H Romer.
\newblock Does monetary policy matter? a new test in the spirit of friedman and schwartz.
\newblock \emph{NBER macroeconomics annual}, 4:\penalty0 121--170, 1989.

\bibitem[Rossi(2013)]{rossi2013advances}
Barbara Rossi.
\newblock Advances in forecasting under instability.
\newblock In \emph{Handbook of economic forecasting}, volume~2, pages 1203--1324. Elsevier, 2013.

\bibitem[Rossi(2021)]{rossi2021forecasting}
Barbara Rossi.
\newblock Forecasting in the presence of instabilities: How we know whether models predict well and how to improve them.
\newblock \emph{Journal of Economic Literature}, 59\penalty0 (4):\penalty0 1135--1190, 2021.

\bibitem[Ryan et~al.(2015)Ryan, Ulrich, Thielen, Teetor, Bronder, and Ulrich]{ryan2015package}
Jeffrey~A Ryan, Joshua~M Ulrich, Wouter Thielen, Paul Teetor, Steve Bronder, and Maintainer Joshua~M Ulrich.
\newblock Package ‘quantmod’.
\newblock 2015.

\bibitem[Sharma et~al.(2016)]{sharma2016forecasting}
Prateek Sharma et~al.
\newblock Forecasting stock market volatility using realized garch model: International evidence.
\newblock \emph{The Quarterly Review of Economics and Finance}, 59:\penalty0 222--230, 2016.

\bibitem[Silver(2016)]{Silver_2016}
Nate Silver.
\newblock 2016 election forecast, Nov 2016.
\newblock URL \url{https://projects.fivethirtyeight.com/2016-election-forecast/}.

\bibitem[Steegen et~al.(2016)Steegen, Tuerlinckx, Gelman, and Vanpaemel]{steegen2016increasing}
Sara Steegen, Francis Tuerlinckx, Andrew Gelman, and Wolf Vanpaemel.
\newblock Increasing transparency through a multiverse analysis.
\newblock \emph{Perspectives on Psychological Science}, 11\penalty0 (5):\penalty0 702--712, 2016.

\bibitem[Sucarrat(2020)]{RePEc:pra:mprapa:100301}
Genaro Sucarrat.
\newblock {garchx: Flexible and Robust GARCH-X Modelling}.
\newblock MPRA Paper 100301, University Library of Munich, Germany, May 2020.
\newblock URL \url{https://ideas.repec.org/p/pra/mprapa/100301.html}.

\bibitem[Timmermann(2006)]{timmermann2006forecast}
Allan Timmermann.
\newblock Forecast combinations.
\newblock \emph{Handbook of economic forecasting}, 1:\penalty0 135--196, 2006.

\bibitem[Tsay(2005)]{tsay2005analysis}
Ruey~S Tsay.
\newblock \emph{Analysis of financial time series}, volume 543.
\newblock John wiley \& sons, 2005.

\bibitem[Wachowicz(2020)]{wachowicz2020wharton}
Erin Wachowicz.
\newblock Wharton research data services (wrds).
\newblock \emph{Journal of Business \& Finance Librarianship}, 25\penalty0 (3-4):\penalty0 184--187, 2020.

\bibitem[Wang et~al.(2020)Wang, Ma, and Liu]{wang2020forecasting}
Lu~Wang, Feng Ma, and Guoshan Liu.
\newblock Forecasting stock volatility in the presence of extreme shocks: Short-term and long-term effects.
\newblock \emph{Journal of Forecasting}, 39\penalty0 (5):\penalty0 797--810, 2020.

\bibitem[Wang et~al.(2023)Wang, Hyndman, Li, and Kang]{wang2023forecast}
Xiaoqian Wang, Rob~J Hyndman, Feng Li, and Yanfei Kang.
\newblock Forecast combinations: An over 50-year review.
\newblock \emph{International Journal of Forecasting}, 39\penalty0 (4):\penalty0 1518--1547, 2023.

\bibitem[Wolfers and Zitzewitz(2016)]{wolfers2016financial}
Justin Wolfers and Eric Zitzewitz.
\newblock What do financial markets think of the 2016 election.
\newblock \emph{Unpublished manuscript. brook. gs/2njI8U3}, 2016.

\bibitem[Zivot(2009)]{zivot2009practical}
Eric Zivot.
\newblock Practical issues in the analysis of univariate garch models.
\newblock In \emph{Handbook of Financial Time Series}, pages 113--155. Springer, 2009.

\end{thebibliography}

\end{document}